\documentclass[aps, prx, notitlepage, twocolumn,longbibliography,superscriptaddress,floatfix]{revtex4-1}
\usepackage{stmaryrd}
\usepackage{amssymb}
\usepackage{amsmath}
\usepackage{hyperref}
\hypersetup{
    colorlinks,
    linkcolor={blue},
    citecolor={blue},
    urlcolor={blue},
    pdfauthor={Jon Nelson, Gregory Bentsen, Steven T. Flammia, Michael J. Gullans},
    pdftitle={Fault-Tolerant Quantum Memory using Low-Depth Random Circuit Codes}
}
\usepackage{listings}

\usepackage[utf8]{inputenc}
\usepackage{color}
\usepackage{qcircuit}
\usepackage{graphicx}
\usepackage{empheq}
\usepackage{placeins}
\usepackage{physics}

\DeclareMathOperator\eye{\mathbb{I}}

\DeclareMathOperator*{\argmax}{arg\,max}

\begin{document}
\title{Fault-Tolerant Quantum Memory using Low-Depth Random Circuit Codes}
\author{Jon Nelson}
\affiliation{Joint Center for Quantum Information and Computer Science, NIST/University of Maryland, College Park, Maryland 20742, USA}
\author{Gregory Bentsen}
\affiliation{Martin A. Fisher School of Physics, Brandeis University, Waltham, Massachusetts 02465, USA}
\author{Steven T. Flammia}
\affiliation{AWS Center for Quantum Computing, Pasadena, California 91125, USA}
\affiliation{IQIM, California Institute of Technology, Pasadena, CA 91125, USA}
\author{Michael J. Gullans}
\affiliation{Joint Center for Quantum Information and Computer Science, NIST/University of Maryland, College Park, Maryland 20742, USA}
\begin{abstract}
    Low-depth random circuit codes possess many desirable properties for quantum error correction but have so far only been analyzed in the code capacity setting where it is assumed that encoding gates and syndrome measurements are noiseless. In this work, we design a fault-tolerant distillation protocol for preparing encoded states of one-dimensional random circuit codes even when all gates and measurements are subject to noise. This is sufficient for fault-tolerant quantum memory since these encoded states can then be used as ancillas for Steane error correction. We show through numerical simulations that our protocol can correct erasure errors up to an error rate of $2\%$. In addition, we also extend results in the code capacity setting by developing a maximum likelihood decoder for depolarizing noise similar to work by Darmawan \textit{et al.}~\cite{darmawan2022low}. As in their work, we formulate the decoding problem as a tensor network contraction and show how to contract the network efficiently by exploiting the low-depth structure. Replacing the tensor network with a so-called ``tropical'' tensor network, we also show how to perform minimum weight decoding. 
    With these decoders, we are able to numerically estimate the depolarizing error threshold of finite-rate random circuit codes and show that this threshold closely matches the hashing bound even when the decoding is sub-optimal.
\end{abstract}

\maketitle

\section{Introduction}
\label{intro}
Random codes have long been a primary object of study in classical coding theory as they are known to achieve the capacity of noisy channels \cite{shannon1948mathematical}. As a result, codes of this type have gained widespread use in practice in the form of random Low-Density Parity Check (LDPC) codes \cite{gallager1962low,mackay1997near,masera2007implementation}. Recent work in quantum information theory suggests that random encoding techniques should enjoy similar success in the quantum setting \cite{gottesman2013fault,kovalev2013fault,brown2013short,brown2015decoupling,fawzi2020constant,gullans2021quantum,leverrier2022quantum}. In particular, the decoupling theorem developed by Hayden \textit{et al.} demonstrates that encoding with Haar random unitaries achieves the quantum channel capacity \cite{hayden2008decoupling}. 

It is also known that the smaller set of random Clifford unitaries generates good (finite rate and linear distance) codes. 
In fact, it has been proven that these codes achieve the quantum Gilbert-Varshamov bound even for $O(\log^3 n)$-depth encoding circuits \cite{brown2013short}. 
More recently, Ref.~
\cite{gullans2021quantum} showed that $O(\log n)$-depth geometrically local circuits in one or higher dimensions are sufficient to encode random codes that approach or saturate, respectively, the capacity of the erasure channel. 
One advantage of these codes is that the rate and error threshold can be easily tuned to reach a desired trade-off. 
This is not the case for many quantum codes where the rate is often a fixed property of the code such as with the toric code \cite{Kitaev_1997}. 
Hypergraph product codes have a similar property in that the rate can be adjusted by changing the underlying classical codes \cite{Tillich_2014,Kovalev_2012}; however, these codes are generally non-local which presents challenges for experimental realization.

These results demonstrate some potential of low-depth random circuit codes for quantum error-correction. 
However, a complete picture of fault tolerance is still missing. 
In the previously mentioned results the encoding circuit and syndrome measurements are assumed to be noiseless. 
This setting is known as the code capacity setting, which only considers one layer of errors in between encoding and decoding for a given code.

In this work, we utilize low-depth random circuit codes in the setting of a fault-tolerant quantum memory. 
Specifically, we outline an error correction protocol that succeeds even when all circuit operations are subject to erasure errors. 
Erasure errors, while not as general as Pauli errors, are nonetheless an important error model to study. They are a helpful benchmark of error correction performance \cite{Delfosse16,Delfosse17}.  Moreover, in many circumstances there exist techniques to convert the dominant source of error into erasure errors. 
For example, such protocols have been shown for quantum hardware based on neutral atoms, trapped ions, and superconducting qubits \cite{Wu_2022,PRXQuantum.4.020358,PhysRevX.13.041022,levine2023demonstrating}. 

To combat these mid-circuit erasure errors, we employ Steane error-correction gadgets \cite{steaneec}, which minimally spread errors while performing syndrome extraction. 
The main prerequisite for Steane error correction is the ability to prepare encoded ancilla states. 
We show that this can be accomplished for low-depth encoding circuits using a distillation protocol composed of bit-flip and phase-flip correction gadgets that closely mirror Steane error correction itself. In addition, this protocol can be executed by 2D geometrically local gates.

The performance of our error correction routine is evaluated through a series of numerical simulations, which indicate that it can sustain erasure errors up to an error rate of approximately $2\%$. 
In our first simulation, we study the entropy of the output state of our distillation protocol. 
The second simulation then uses these states to execute the full Steane error correction gadget. 
Performance is evaluated by maximally entangling the data block with a reference and studying the quantum mutual information between the two after applying many rounds of error correction. 
In the final simulation, Pauli errors are randomly sampled throughout the circuit and the spacetime code framework of \cite{bfhs, delfosse2023spacetime} is used to decode the errors. 
In each case the scaling behavior is studied by varying the encoding depth $d$ and the number of rounds of distillation $q$. 
Our results consistently indicate threshold behavior at an error level of roughly $p = 0.02$. 
This result shows promising evidence that low-depth random circuit codes can be used for fault-tolerant quantum memory. 

Similar results have previously been achieved for a wide variety of other codes including the surface code \cite{Kitaev_2003,Dennis_2002,Fowler2009}, LDPC codes \cite{Tremblay_2022,bravyi2023highthreshold,xu2023constantoverhead}, hyperbolic Floquet codes \cite{higgott2023constructions,fahimniya2023faulttolerant}, and others. In addition there has also been exciting progress on the experimental realization of these fault-tolerant schemes \cite{wang2023faulttolerant,ryananderson2022implementing,gupta2023encoding,Krinner_2022,acharya2022suppressing}.
Our particular approach shows that a high threshold can be achieved at the cost of an overhead that depends exponentially on the encoding depth. As shown in Ref.~\cite{gullans2021quantum}, it is sufficient for the encoding depth to scale logarithmically in the number of qubits, and so the exponential overhead with respect to the depth is in turn polynomially large in terms of the number of qubits.

In addition to developing the fault-tolerant picture, we also present practical tools for decoding low-depth random circuit codes in the code capacity setting. 
Whereas decoding random stabilizer codes is hard in general (optimal decoding is \#P-Complete for arbitrary quantum stabilizer codes \cite{iyer2015hardness}), low-depth random circuit codes naturally possess a quasi-local structure that can be exploited for efficient decoding.
Utilizing this local structure, we present two different polynomial-time decoding algorithms for finite-rate, low-depth random-circuit codes. 
The first of these is nearly optimal in the sense that it determines the most likely error on each logical qubit while marginalizing over the rest as is also the case in Ref.~\cite{darmawan2022low}. 
This is closely related to maximum likelihood decoding, which finds the most likely error over all logicals at once. 
The second decoder we present is a minimum weight decoder, meaning it finds the minimum weight error that is consistent with the syndrome measurement. 
Minimum weight decoders of this type greatly simplify the decoding problem and can lead to faster decoding implementations \cite{Dennis_2002,duclos2010fast}. 
Both decoding algorithms have runtimes that scale exponentially with circuit depth, which is still efficient when the depth grows no faster than logarithmically in the number of qubits.

Armed with these decoding algorithms, we show that low-depth random-circuit codes maintain high performance against the depolarizing channel, which is a more challenging setting than the previously studied erasure channel. 
More specifically, we show that the threshold of this family of codes closely matches
the hashing bound, which supports the results of Darmawan \textit{et al.} \cite{darmawan2022low}. 
We also employ the sub-optimal minimum weight decoder and show that its performance is comparable to the optimal decoder across a range of encoding rates. 
Our results are complementary to tensor-network decoding algorithms appearing in Ref.~\cite{darmawan2022low}, which were developed independently and contemporaneously with the algorithms presented here. 
While the maximum-likelihood decoder shares many similarities with the tensor-network decoder of Darmawan \textit{et al.}, the minimum weight decoder instead utilizes tropical algebra and builds on recent developments in numerical spin glass techniques \cite{liu2021tropical}. 
This tropical tensor network decoder allows us to demonstrate a high threshold even in the setting of sub-optimal decoding. 
Although the algorithms were developed independently, we adopt the same setup as Ref.~\cite{darmawan2022low} for our code capacity numerical simulations.

A primary technique that we utilize throughout is a well-known mapping between stabilizer codes and classical Ising spin models~\cite{Dennis_2002, chubb2021statistical}. 
In this framework, decoding is equivalent to minimizing the free energy of the spin model. 
When the temperature of the model is set according to the Nishimori conditions \cite{nishimori1981internal}, this minimization problem corresponds to maximum likelihood decoding. 
Alternatively, when the temperature approaches zero, minimizing the free energy is equivalent to finding the minimum weight error \cite{chubb2021statistical}. 

The paper is organized as follows: we begin in Section \ref{codes} by describing the specific class of low-depth random-circuit codes considered in this work. Next, in Section \ref{decoding} we formally define the quantum decoding problem and review the mapping from stabilizer codes to classical Ising spin models. In Section \ref{channelcoding}, we present our results for estimating the depolarizing error threshold in the code capacity setting. Then, in Section \ref{ftprotocol}, we present our fault-tolerant distillation protocol. Finally, we present three numerical simulations for estimating the fault-tolerant threshold of our protocol in Section \ref{ftthreshold} and conclude with a brief discussion of future work in Section \ref{discuss}.

\begin{figure}
    \centering
    \includegraphics[width=0.95\columnwidth]{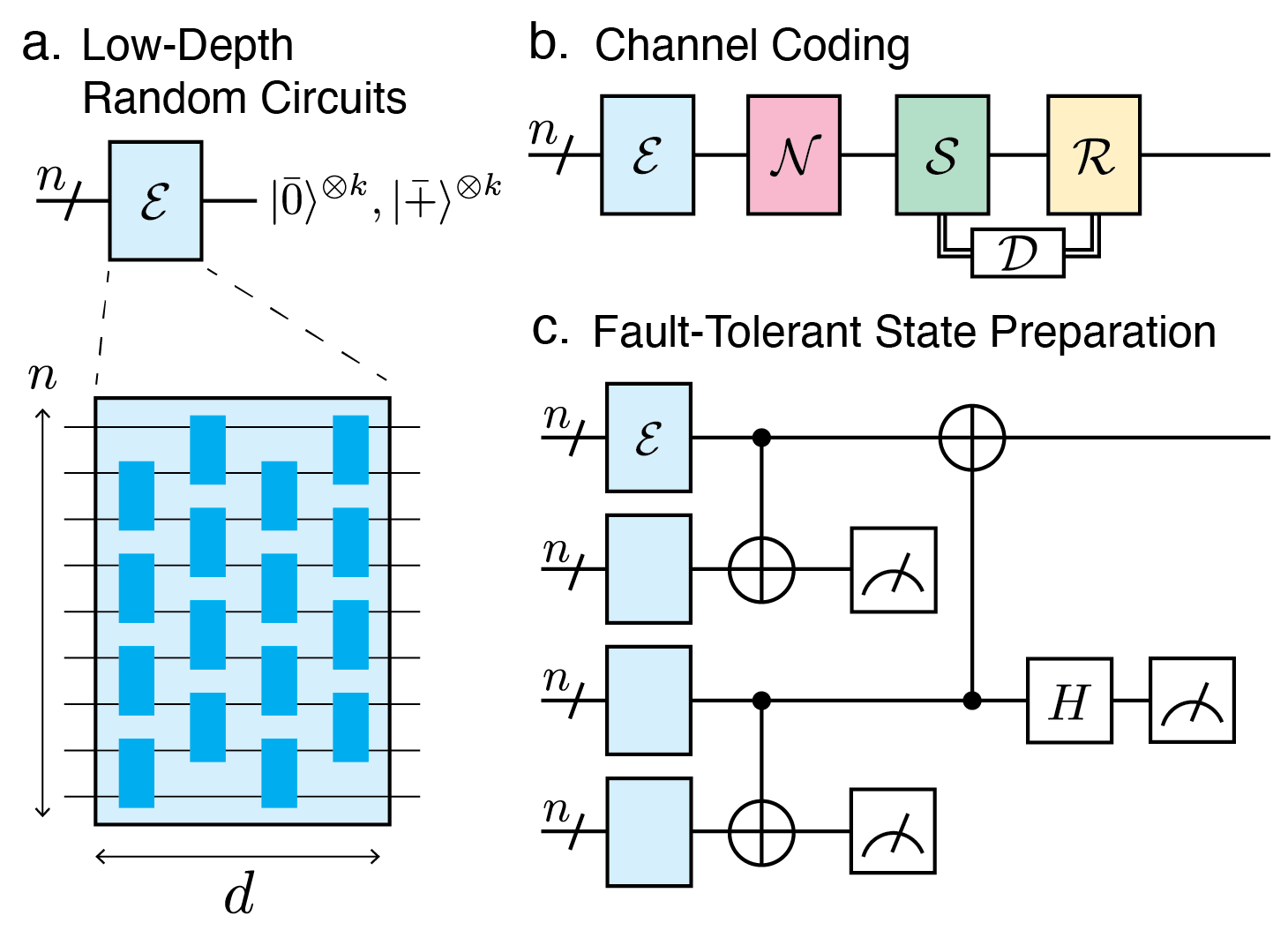}
    \caption{Low-depth random circuit codes for fault-tolerant quantum memory. In this work we study brickwork encoding circuits $\mathcal{E}$ (a) of low depth $d$, where blue rectangles represent randomly sampled two-qubit Clifford gates. We first analyze these codes in the code capacity setting (b), where a noiseless encoding step is followed by a noisy channel $\mathcal{N}$, a noiseless syndrome measurement $\mathcal{S}$, and a noiseless recovery channel $\mathcal{R}$. Then we describe a fault-tolerant protocol for preparing encoded states (c) where noise occurs anywhere throughout the circuit.}
    \label{fig:lowdepthfaulttolerance}
\end{figure}

\section{Low-Depth Random Circuits}
\label{codes}
We consider 1D encoding circuits $\mathcal{E}$ of depth $d$ acting on a system of $n$ qubits, where randomly sampled 2-qubit Clifford gates are applied to neighboring qubits in a brickwork arrangement as illustrated in Fig. \ref{fig:lowdepthfaulttolerance}.a. Note that open boundary conditions are used to simplify the decoding later on. Since these are Clifford circuits, they can be used as the encoding circuit of a stabilizer code in the following way. First, each input qubit to the encoding circuit is either associated with a stabilizer generator or logical qubit of the code. The input qubits associated with stabilizer generators are set to the state $\ket{0}$. Since this is stabilized by $Z_i$, the output state of the circuit will be stabilized by $UZ_iU^\dagger$ where $U$ is the Clifford circuit. Therefore, $UZ_iU^\dagger$ is considered a stabilizer generator of the code. The remainder of the input qubits correspond to logical qubits of the code. For example, if the rate of the code is $k/n = 1/2$, then every other input qubit is associated to a stabilizer generator and thus set to the state $\ket{0}$, and the rest are associated to logical qubits. To prepare the encoded all-zero state for this code, $\ket{\bar{0}}^{\otimes k}$, the input qubits associated with logical qubits are also set to $\ket{0}$ so that the output state of the circuit is now stabilized by $UZ_iU^\dagger$ for all qubits $i$. These operators $UZ_iU^\dagger$ for input qubits $i$ associated with logical qubits make up the logical $Z$ operators of the code. Similarly, $UX_iU^\dagger$ for these same input qubits make up the logical $X$ operators of the code.

Since each layer is made up of only 2-qubit gates, the stabilizer and logical generators of the code can only grow by one qubit on either side per layer. 
Therefore, the overall support of the generators is at most $2d$, which will become crucial when we later describe our decoding algorithms.

One issue introduced by the open boundary conditions is that logical qubits near the boundary become stunted. This is most prevalent when the depth $d$ is much larger than the inverse of the rate since the outermost logical will be a distance $O(n/k)$ away from the boundary but would normally have support up to $d$ qubits from its center. In order to remedy this, we pad the boundary with extra stabilizer generators as done in Ref.~\cite{darmawan2022low}. In particular, approximately $2d$ stabilizers are added to either side of the circuit to ensure that no logical qubit is within $2d$ qubits of the boundary. This padding becomes negligible when the number of qubits $n$ is large relative to $d$ as in our case.

When extending our results to the fault-tolerant setting, we also consider a related class of circuits that we call random CSS circuits. Recall that Calderbank-Shor-Steane (CSS) codes are stabilizer codes where the stabilizers can be chosen to be either an $X$-type or $Z$-type generator. An $X$-type ($Z$-type) Pauli operator only contains identities and $X$'s ($Z$'s). Random CSS circuits are thus Clifford circuits that maintain the CSS property of the associated code. More specifically, each gate in the circuit is randomly sampled from the subset of Clifford gates that map $X$-type Pauli's to $X$-type Pauli's and $Z$-type to $Z$-type. Any input qubit starting in the state $\ket{+}$ is stabilized by $X_i$ and thus becomes $UX_iU^\dagger$ after applying the encoding circuit. This must also be an $X$-type stabilizer since gates preserve the stabilizer type. Similarly, any input qubit starting in the state $\ket{0}$ is stabilized by $Z_i$ and becomes $UZ_iU^\dagger$ which is a $Z$-type stabilizer. Finally, as in the non-CSS case, logical $X$ and $Z$ operators are defined by $U X_i U^\dagger$ and $U Z_i U^\dagger$, respectively, where $i$ is the index of an input qubit associated with a logical operator rather than a stabilizer generator. For example to encode the all-zero state, $\ket{\bar{0}}^{\otimes k}$, one would first set the input qubits associated with $X$-type stabilizers to $\ket{+}$ and $Z$-type stabilizers to $\ket{0}$. Since the rest of the qubits are associated to the logical zero state they would all be initialized to $\ket{0}$. Finally, the random CSS circuit is applied to this input state to get the encoded all-zero state. Note that we use periodic rather than open boundary conditions for these random CSS circuits since we analyze them subject to erasure errors, which do not have the same decoding difficulties as Pauli errors.

CSS codes have many convenient properties for designing fault-tolerant protocols. In fact, Steane error correction only works for CSS codes because logical CNOT gates are transversal for any CSS code, which is not true in general for any stabilizer code. We will leverage this property in our protocol and so these CSS circuits are crucial for implementing random circuit codes in the fault-tolerant setting.

\section{Maximum Likelihood Decoding}
\label{decoding}
Given a stabilizer code defined by its stabilizer group $S$ and logical group $L$, any Pauli error $E$ can be decomposed as
\begin{equation}
    E = S_a L_{\ell} C_s
\end{equation}
where $S_a \in S$ is a stabilizer element, $L_{\ell} \in L$ is a logical operator, and $C_s$ is the canonical error associated with the syndrome $s$. Recall that this syndrome is obtained by measuring each of the stabilizer generators and that $C_s$ can be efficiently calculated using the stabilizer tableau framework and Gaussian elimination \cite{Aaronson_2004}. Since multiplying by a member of the stabilizer group has no effect on the codespace, all errors that differ only by a stabilizer element are treated as equivalent. We therefore define the logical equivalency class $\overline{L_{\ell}}$ as the collection of errors that differ from the logical operator $L_{\ell}$ by only a stabilizer element:
\begin{equation}
    \overline{L_{\ell}} = \{S_a L_{\ell} | S_a \in S\}.
\end{equation}
The goal of maximum likelihood decoding is to determine which logical class $\overline{L_{\ell}}$ is most likely given the error syndrome $s$. This can be expressed as the following optimization problem: 
\begin{equation}
    \label{mld}
    \argmax_{\ell} \sum_{E \in \overline{L_{\ell}}} P(E) = \argmax_{\ell}\sum_{S_a \in S} P(S_a L_l C_s)
\end{equation}
where $P(E)$ is the probability that error $E$ occurs under the given error channel and observed syndrome measurements. The error can then be corrected by applying the operator $L_{\ell}C_s$. Multiplying by $L_{\ell}$ cancels the logical operation of the error and multiplying by $C_s$ returns it to the codespace.

To solve this decoding problem, we employ a well-known mapping between stabilizer codes and classical spin models \cite{Dennis_2002,chubb2021statistical}. The main idea is to view the probabilities $P(E)$ as Gibbs weights $e^{- \beta H_E(\vec{s})}$ in a statistical mechanical model of Ising spins $\vec{s}$ where $s_i = \pm 1$. Each spin $s_i$ is associated with a stabilizer generator $S_i$.
Given a stabilizer code with stabilizer group $S = \langle S_i \rangle$ on $n$ physical qubits and a Pauli error $E \in \mathcal{P}^{\otimes n}$, the classical spin Hamiltonian is defined as
\begin{equation}
    H_E(\vec{s}) = \sum_{\sigma} J \llbracket E, \sigma \rrbracket \prod_{\llbracket \sigma, S_i \rrbracket = -1} s_i,
\end{equation}
where the sum is over all single-weight Pauli operators $\sigma$. The scalar commutator is defined as $\llbracket A, B \rrbracket = \mathrm{Tr}[A B A^{-1} B^{-1}]/2^n$ and is equal to $1$ if $A$ and $B$ commute and $-1$ if they anti-commute. This fully describes the commutativity relations of Pauli operators since they always either commute or anti-commute. 

For depolarizing noise with $X, Y, Z$ errors occurring with probability $p/3$ each, we fix the inverse temperature $\beta$ by demanding that the model satisfies the Nishimori conditions:
\begin{equation}
\beta J = - \frac{1}{4} \log \frac{3(1-p)}{p}.
\end{equation} 

For precisely this relationship between inverse temperature and coupling strengths, the Gibbs weight in the stat-mech model is exactly proportional to the error probability:
\begin{equation}
e^{-\beta H_E(\vec{1})} \propto P(E)
\end{equation}
where $\vec{1}$ is the configuration with all spins pointing up ($s_i = +1$ for all $i$).
Furthermore, flipping spin $s_i$ from $+1$ to $-1$ corresponds to multiplying the error $E$ by the corresponding stabilizer generator $S_i$. Also, note that every stabilizer element can be produced by multiplying together some subset of the stabilizer generators. Thus, summing over every possible configuration of the spin variables is equivalent to summing over all stabilizer elements multiplied by $E$. This leads to the following interpretation of the partition function:
\begin{equation}
Z_E = \sum_{\vec{s}} e^{-\beta H_E(\vec{s})} \propto \sum_{S_a \in S}P(S_a E)
\end{equation}

Notice that if $E = L_{\ell}C_s$ then $Z_E$ is proportional to the probability of the logical class $\ell$ given the syndrome $s$ from Eq. \ref{mld}. In Appendix \ref{app:tensor}, we give an efficient method for calculating $Z_E$ for low-depth circuit-codes, and so this immediately suggests a maximum likelihood decoder for zero-rate codes. Namely, calculate $Z_E$ where $E = L_{\ell}C_s$ for each logical class $l$ and choose the $l$ that corresponds to the largest partition function. For finite-rate codes this becomes more challenging since the number of logical classes grows exponentially in the number of logical qubits. To address this, we show in Appendix \ref{app:marginaldecoding} how to add logical classes as spins in the stat-mech model to decode logical qubits individually.

While the maximum-likelihood decoder described above provides a near-optimal decoding algorithm, this is often not necessary for good performance. In fact, it is often more practical to sacrifice optimality for efficiency. A common simplification of the decoding problem is to find the minimum weight Pauli operator that is consistent with the error syndrome. This is called minimum weight decoding and is often easier since it ignores the fact that many different errors can be logically equivalent. We describe minimum weight decoding in more detail in Appendix \ref{app:minweightdecoding}.

\section{Code Capacity Threshold}
\label{channelcoding}

Equipped with our encoding circuits $\mathcal{E}$ and a set of algorithms for decoding, we are now ready to evaluate the performance of these random circuit codes in the code capacity setting. To perform the threshold estimation, we adopt the protocol of Ref.~\cite{darmawan2022low}. 
In this numerical simulation, the number of physical qubits is held constant at $n=50$ and only the depth is varied. For each depth ranging from $d=4$ to $d=7$, 10,000 different random codes and Pauli errors are sampled. 
The errors are decoded using both the marginal and minimum weight decoder and the failure rate $p_{m}$ of each logical qubit $m$ is calculated individually. 
The average failure rate over all logical qubits is denoted by $\bar{p}_L = 1/k \sum_{m} p_{m}$ \footnote{Note that $\bar{p}_L$ is similar to $p'_L$ in \cite{darmawan2022low} in that they both consider failure probabilities of individual logical qubits.} and plotted with respect to various error rates. 
This is repeated for encoding rates of $1/3,1/4,1/5$, $1/10$, and $1/50$. These plots are displayed in Figure \ref{fig:app_marg} and \ref{fig:app_mw} of Appendix \ref{app:supplresults}.

\begin{figure}
\centering
\includegraphics[width=\linewidth]{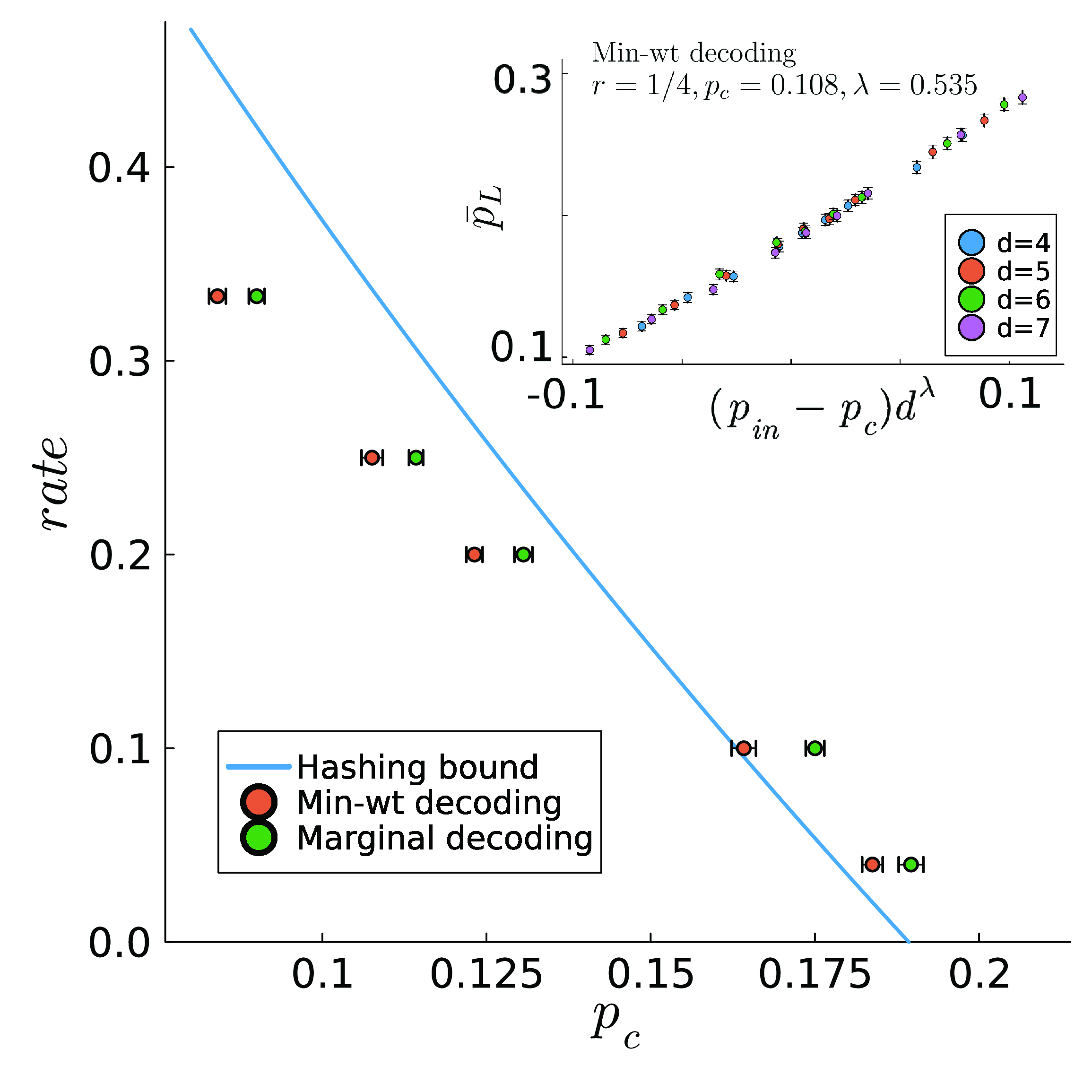}
\caption{Threshold estimation for random-circuit codes of various encoding rates. Codes are generated on systems of $n = 50$ qubits (excluding the boundary padding) using random stabilizer circuits of depths $d = 4,5,6,7$. Error thresholds for both marginal decoding and minimum weight decoding are presented. For each datapoint, the threshold is estimated by fitting our finite-size scaling ansatz. Jackknife resampling is used to estimate the standard deviation of our fitting procedure which is shown as error bars. An example of this scaling collapse is shown in the inset for rate = $1/4$. The rest of the crossing plots and associated collapses can be found in Figures \ref{fig:app_marg} and \ref{fig:app_mw} of Appendix \ref{app:supplresults}.}
\label{fig:rate4mwd}
\end{figure}

It can be seen in these figures that at the critical error rate there is a crossing point. 
Below this critical error rate, increasing the depth exponentially suppresses the average failure rate $\bar{p}_L$. 
This is closely related to the code's error threshold, which is defined instead in terms of the probability that \emph{any} logical qubit fails,  denoted as $p_L$. The authors of \cite{darmawan2022low} argue that this crossing point in $\bar{p}_L$ does indeed correspond to a crossing in $p_L$ due to the observation by \cite{gullans2021quantum} that logical qubits have short-ranged correlations. It is helpful in this case to study $\bar{p}_L$ instead of $p_L$ directly since, as pointed out by \cite{darmawan2022low}, the logical qubits near the boundary behave differently than logical qubits in the bulk. Averaging over the failure rate of all logical qubits can help avoid these boundary effects.

To accurately estimate this threshold, a finite-size scaling collapse of the data is performed \cite{Wang_2003,Harrington2004AnalysisOQ,PhysRevA.81.022317} using the following ansatz:
\begin{equation}
\label{eq:ansatz}
    \bar{p}_L = f( d^{\lambda}(p-p_c)).
\end{equation}
$f(x)$ is approximated as $f(x) = A + Bx + Cx^2$ by taking the Taylor expansion around $x=0$, which corresponds to $p=p_c$. This introduces a total of five fitting parameters: $p_c$, $\lambda$, $A$, $B$, and $C$. These parameters are estimated by minimizing the mean-squared error of our model with the observed data. In order to estimate the standard deviation of this fitting procedure, jackknife resampling is conducted and this is represented as error bars in Figure \ref{fig:rate4mwd}. It is also worth noting that the specific range of $p_{in}$ that is used in the fitting procedure is important. Values of $p_{in}$ close to $p_c$ are more accurately approximated by the Taylor expansion but are not as helpful for estimating the fit parameters. In order to eliminate outliers where the Taylor expansion is no longer accurate, the range of $p_{in}$ is truncated to exclude data points that significantly increase the standard deviation of the fit. 

This finite-size scaling collapse is presented for rate = $1/4$ and minimum weight decoding in the inset of Figure \ref{fig:rate4mwd}. The scaling collapses for the rest of the encoding rates are included in Figures \ref{fig:app_marg} and \ref{fig:app_mw} of Appendix \ref{app:supplresults}. It can be seen that the estimated $p_c$ leads to an approximately linear relationship between the scaling ansatz and $\bar{p}_L$ regardless of the depth $d$, suggesting the presence of a threshold at this value. The estimated value for $p_c$ at each encoding rate is plotted in Figure \ref{fig:rate4mwd} along with the standard deviation estimated through jackknife resampling.

A key observation from these results is that the critical error rate remains close to the hashing bound even when using the minimum weight decoder. 
This suggests that it is sufficient to solve the easier minimum weight decoding problem to achieve near-optimal performance. 
One reason decoding in the quantum setting is often harder than the classical setting is that errors are highly degenerate, meaning one logical class contains many equivalent Pauli operators. 
However, in the minimum weight decoding problem, logical classes are completely ignored and the problem is reduced to the classical case of finding the most likely error rather than most likely logical class. 
Our results show that this simplification comes at only a small cost in the threshold and we hope to take advantage of this by exploring approximate decoding methods in future work for solving this problem more efficiently.

\section{Fault-Tolerant State Preparation and Error Correction}
\label{ftprotocol}
\begin{figure}[htpb]
\mbox{
\Qcircuit @C=1em @R=.7em 
{\lstick{} & \ctrl{1} & \targ & \qw & \qw\\
\lstick{\ket{\bar{+}}^{\otimes k}} & \targ & \qw & \meter \\
\lstick{\ket{\bar{0}}^{\otimes k}} & \qw & \ctrl{-2} & \gate{H} & \meter
}}
\caption{Steane error correction gadget for measuring stabilizers of a CSS code. Note that each wire represents a code-block of $n$ qubits and that the CNOT and Hadamard operations are implemented transversally on the entire code-block.}
\label{fig:steaneec}
\end{figure}
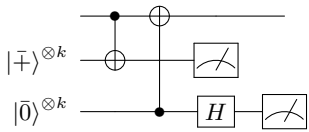

In the previous section, we assumed that encoding $\mathcal{E}$, syndrome measurement $\mathcal{S}$, and recovery $\mathcal{R}$ are all implemented perfectly. In practice these operations will be implemented on noisy hardware and so a more comprehensive error model must be studied in order to use these codes in practice. To address this, we next consider a method for executing error correction even when errors can occur at every time step of the circuit. We use the Steane error correction circuit, as shown in Figure \ref{fig:steaneec}, to perform syndrome extraction. Steane error correction controls the spread of errors by using transversal gates, which propagate Pauli errors to at most one other qubit. Since only CSS codes are guaranteed to have transversal CNOT gates, it is important that we use CSS encoding circuits for this scheme to work. In addition, this circuit relies on the ability to prepare the encoded all-zero state, $\ket{\bar{0}}^{\otimes k}$, and all-plus state, $\ket{\bar{+}}^{\otimes k}$, which can be a highly nontrivial task. To this end, we design a distillation procedure to prepare these states in a way that is robust to mid-circuit errors. We note that similar methods have also been proposed for preparing encoded states of concatenated codes \cite{Chamberland_2017}.

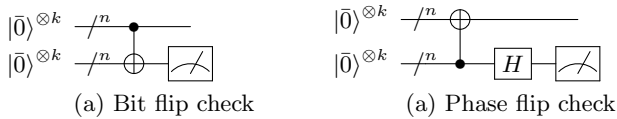
\begin{figure}[t!]

\begin{minipage}[b]{0.45\columnwidth}
\centering
\mbox{
\Qcircuit @C=1em @R=.7em 
{\lstick{\ket{\bar{0}}^{\otimes k}} &  {/^n} \qw & \ctrl{1} & \qw \\
\lstick{\ket{\bar{0}}^{\otimes k}} &  {/^n} \qw & \targ & \meter
}}
\newline
\newline
(a) Bit flip check
\end{minipage}
\hspace{0.5cm}
\begin{minipage}[b]{0.45\columnwidth}
\centering
\mbox{
\Qcircuit @C=1em @R=.7em 
{\lstick{\ket{\bar{0}}^{\otimes k}} &  {/^n} \qw & \targ & \qw & \qw \\
\lstick{\ket{\bar{0}}^{\otimes k}} &  {/^n} \qw & \ctrl{-1} & \gate{H} & \meter
}}
\newline
\newline
(a) Phase flip check
\end{minipage}
\caption{Bit and phase correction circuits. Note that $\ket{\bar{0}}^{\otimes k}$ represents $k$ logical qubits in the all-zero state. Each wire in the diagram represents an entire code-block made up of $n$ physical qubits. In addition, the CNOT represents the transversal CNOT applied across the two code-blocks and the Hadamards and measurements are applied to each qubit in the block. Although not depicted here, erasure errors can occur at any time step in the circuit. Note that single-qubit gates are absorbed into adjacent 2-qubit gates when determining what constitutes a time step. Finally, the same circuit can also be used for correcting $\ket{\bar{+}}^{\otimes k}$.}
\label{fig:cnotcheck}
\end{figure}

First, a random CSS circuit code, as defined in Section \ref{codes}, is sampled and many different noisy copies of $\ket{\bar{0}}^{\otimes k}$ or $\ket{\bar{+}}^{\otimes k}$ are prepared depending on which is the desired output state. These copies are then paired up and checked against each other using the circuits in Figure \ref{fig:cnotcheck}, which either check whether a bit or phase error occurred. 

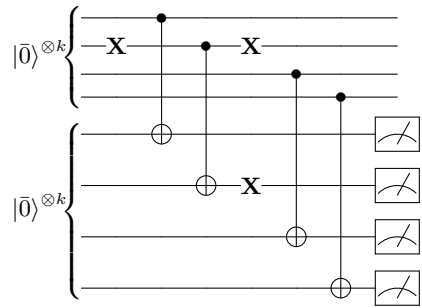
\begin{figure}[htpb]
\centering
\mbox{
\Qcircuit @C=1em @R=.7em 
{\lstick{} & \qw & \ctrl{4} & \qw & \qw & \qw & \qw & \qw \\
\lstick{}  & \push{\textbf{X}} \qw & \qw & \ctrl{4} & \push{\textbf{X}} \qw & \qw & \qw & \qw \\
\lstick{} & \qw & \qw & \qw & \qw  & \ctrl{4}  & \qw & \qw \\
\lstick{} & \qw & \qw & \qw & \qw & \qw & \ctrl{4} & \qw 
\inputgroupv{1}{4}{.8em}{1.5em}{\ket{\bar{0}}^{\otimes k}} \\
\lstick{} & \qw & \targ & \qw  & \qw & \qw & \qw & \meter \\
\lstick{} & \qw & \qw  & \targ & \push{\textbf{X}} \qw & \qw & \qw & \meter \\
\lstick{} & \qw & \qw & \qw & \qw & \targ & \qw & \meter \\
\lstick{} & \qw & \qw  & \qw & \qw & \qw & \targ & \meter 
\inputgroupv{5}{8}{.8em}{3em}{\ket{\bar{0}}^{\otimes k}} \\
}}

\caption{Example of an $X$ error being detected by our bit flip correction circuit. First, the CNOT gate propagates the $X$ error from the data block to the ancilla block. On the ancilla block, the $X$ error then flips the corresponding measurement outcome. Since the measurements are performed on a CSS codeword, they have a particular form as discussed in the text. The measurement error can thus be detected by comparing to the expected measurement given no errors. The phase flip correction circuit detects $Z$ errors in a similar manner.}
\label{fig:errorprop}
\end{figure}

In the case of bit flip correction, the CNOT gates in this circuit will propagate $X$ errors (bit flips) from the first ancilla block to the second. This is depicted in Figure \ref{fig:errorprop}. The $X$ errors on the second ancilla block will then flip the corresponding measurement outcome. By inspecting which measurements have been flipped, it is possible to detect which qubits may have experienced an $X$ error.

More concretely, when there are no errors, the transversal CNOT gates perform the identity since these act as logical CNOTs where the controls are the encoded zero states. In this case, the circuit simply measures the second ancilla block in the standard basis. Recall that codewords of CSS codes in the computational basis are superpositions over codewords of a classical linear code defined by the $Z$-type stabilizers. Therefore, the measurement result will sample one of these classical codewords with equal probability. Since $X$ errors cause bit flips on the measured codeword, correcting $X$ errors is equivalent to the classical decoding problem where the noisy codeword is the measurement outcome. Although we focused on $X$ errors here, phase flip correction works in the same way.

After applying these correction circuits, the measurement destroys one of the two states and the process is then repeated for the remaining half of the states until there is only one state left. In each round of this distillation procedure, the comparison circuit alternates between either correcting for bit flips or phase flips. 

\begin{figure}[t!]
\mbox{
\Qcircuit @C=1em @R=.7em 
{\lstick{\ket{\bar{0}}^{\otimes k}} & \ctrl{1} & \qw & \targ & \qw & \qw & \ctrl{4} & \qw & \targ & \qw\\
\lstick{\ket{\bar{0}}^{\otimes k}} & \targ & \meter \\
\lstick{\ket{\bar{0}}^{\otimes k}} & \ctrl{1} & \qw & \ctrl{-2} & \gate{H} & \meter\\
\lstick{\ket{\bar{0}}^{\otimes k}} & \targ & \meter \\
\lstick{\ket{\bar{0}}^{\otimes k}} & \ctrl{1} & \qw & \targ & \qw & \qw & \targ & \meter\\
\lstick{\ket{\bar{0}}^{\otimes k}} & \targ & \meter \\
\lstick{\ket{\bar{0}}^{\otimes k}} & \ctrl{1} & \qw & \ctrl{-2} & \gate{H} & \meter\\
\lstick{\ket{\bar{0}}^{\otimes k}} & \targ & \meter \\
\vdots & & & & & & & & \ctrl{-8} & \qw \\
\\
}}
\caption{Partial circuit for our distillation procedure. In this case, the first round contains only bit flip checks and the type of check alternates in each subsequent round. The top data block will be the outputted encoded state. 
}
\label{fig:steaneverificationcircuit}
\end{figure}
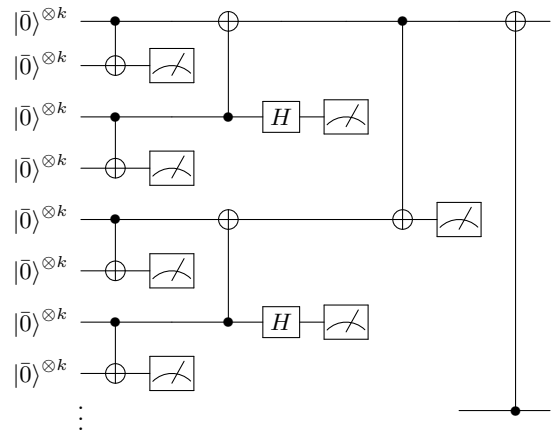

Although the number of ancilla states in this protocol grows exponentially with the number of rounds of distillation, our simulations show that the number of necessary rounds, which we denote as $q$, scales linearly with the encoding depth $d$ which is logarithmic in the number of qubits. This results in only a polynomial overhead in ancilla qubits. To make the procedure explicit, Figure \ref{fig:steaneverificationcircuit} depicts three rounds of the distillation procedure.

The output of this distillation procedure is either an encoded all-zero or all-plus state with potentially some residual errors. The measurements from the bit and phase flip checks can be used to identify what errors may have occurred. Rather than correcting these errors explicitly, it is sufficient to keep track of the errors and adjust the Pauli frame accordingly. The goal is that after many rounds of this distillation procedure, all errors will have been identified up to a stabilizer element. These encoded states can then be used for syndrome measurement by performing the Steane error correction gadget in Figure \ref{fig:steaneec}. Note that although the error locations are not depicted in this figure, all gates and measurements are considered to be noisy.

This circuit can be executed by 2D geometrically local gates by first preparing each noisy encoded state on a column of qubits. With this configuration, neighboring columns are paired up so that the transversal CNOT between code-blocks is performed by CNOT gates acting on neighboring qubits. After each round of bit or phase flip checks, the non-measured code-block must be shifted forward so that it once again neighbors another encoded state. This shift can be performed by SWAP gates on neighboring columns of qubits. 
After distillation, the Steane error correction gadget can be performed in a similar fashion so that the entire protocol is implementable in 2D. In our simulations we did not restrict to 2D circuits, but since the only difference is one round of swap gates in between each distillation round, we believe this has a minimal effect on our results.

\section{Fault-Tolerant Threshold}
\label{ftthreshold}
To evaluate the error threshold of this protocol, we performed a variety of numerical simulations. In each case, errors are taken to be erasure errors, which for a single qubit is defined by the following channel:
\begin{align}
    \mathcal{N}(\rho) = (1-p) \rho \otimes \ketbra{0}{0} + p \eye / 2 \otimes \ketbra{1}{1}
\end{align}
In other words, with probability $p$ the qubit is replaced with the maximally mixed state and the location of the error is signalled by the $\ket{1}$ state of the auxiliary register. 
This is equivalent to the depolarizing channel except with the added benefit of knowing the locations of where each error occurred. 
This equivalence is a key reason why general Pauli errors can often be converted to erasure errors in physical architectures.
In our simulations, there is a single error rate $p$ and for every time step an erasure error occurs at each qubit with probability $p$.

\subsection{Entropy density of prepared states}
\label{subSection:entropydensity}
In the first numerical simulation, erasure errors are randomly sampled at each time step of the distillation protocol and the state's entropy density is calculated at the end of every other round. Erasure errors are simulated by preparing ancilla qubits in the maximally mixed state, swapping these ancillas with the qubits being erased, and tracing out the old qubits. This circuit can be simulated using an extension of Gottesman-Knill simulation that allows for mixed stabilizer states \cite{Aaronson_2004}. This is because the entire protocol only contains Clifford circuits, Pauli measurements, and maximally mixed ancilla qubits, which can be represented as mixtures over stabilizer states. The entropy is calculated using the formula: $S(\rho) = n-r$ \cite{PhysRevX.7.031016,PhysRevA.71.022315,HAMMA200522} where $n$ is the number of physical qubits and $r$ is the stabilizer rank of the state (the size of the minimal generating set of the state's stabilizer group).

Intuitively, the entropy density of the encoded state reflects how much information is gained about the circuit errors from the measurements in the distillation procedure. If the error can be exactly identified up to multiplication by some stabilizer of the encoded state (here, logical $Z$ operators are considered “stabilizers" of the state $\ket{\bar{0}}^{\otimes k}$ ) then the resulting state will have zero entropy density. After adjusting the Pauli frame to account for the known error that occurred, this state can be used as a noiseless ancilla state for Steane error correction. However, if there is still uncertainty over which error occurred, then the resulting state will be a mixture over the different possible errors applied to the state.

\begin{figure}[t!]
    \centering
    \includegraphics[width=0.95\columnwidth]{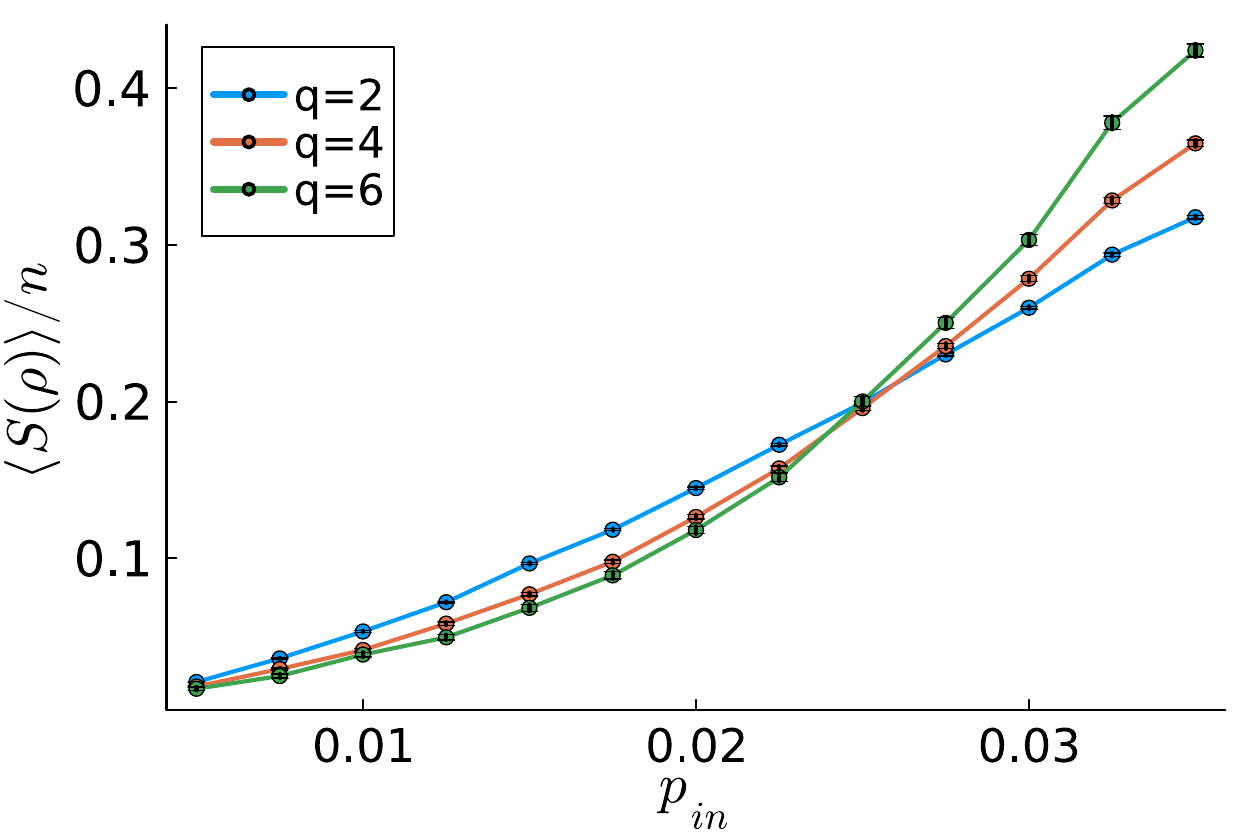}
    \caption{Entropy density of the prepared encoded state after various rounds of our distillation protocol. $n=51$ physical qubits are used to encode $k=17$ logical qubits which is a rate of $1/3$. In each case, $d=6$ layers of a randomly sampled CSS circuit are used as the encoding circuit. $q=2,4,6$ rounds of distillation are performed. The experiment is repeated $1000$ times and the standard error is displayed as error bars.}
    \label{fig:entropydensity}
\end{figure}

At each time step of the protocol (including each layer of the encoding circuit and each round of distillation) erasure errors occur with probability $p$ on each qubit. For each error rate $p$, 1000 different CSS encoding circuits and erasure error patterns are randomly sampled and simulated. The entropy density is calculated after every two rounds of distillation (one round of bit flip correction and one round of phase flip correction). The average entropy density, $\langle S(\rho) \rangle /n$, and standard error are plotted as a function of $p$ in Figure \ref{fig:entropydensity} for an encoding rate of $1/3$. To prepare the state $\ket{\bar{0}}^{\otimes k}$, the input state to the encoding circuit is initialized as $\ket{+00+00+00 \dots}$ where the first qubit of the three qubit pattern is associated with an $X$-type stabilizer, the second qubit is associated with a logical qubit in the state $\ket{\bar{0}}$, and the third qubit is associated with a $Z$-type stabilizer. 

In this plot, there is a clear crossing point near $p=2.5\%$. Below this point increasing the number of rounds of distillation improves the purity of the output state. On the other hand, the distillation procedure only amplifies the noise when the error rate is above $2.5\%$. This is evidence of an erasure threshold for our state preparation procedure. 

Notice that this threshold is only for the state preparation routine and not for the error correction scheme as a whole. These thresholds should not be expected to align since certain states may be easier to prepare but may not be as useful for Steane error correction. For instance, the unencoded all-zero and all-plus state are both easy to prepare but provide no useful information when used as the ancillas in Steane error correction. Therefore, it is reasonable that in this particular case the state preparation threshold is a bit higher than our reported threshold of $2\%$ for the entire protocol.

\subsection{Evaluating Steane error correction gadget using entanglement test}

The next step is to test the full Steane error correction gadget, which uses two encoded ancillas prepared by our distillation procedure. In order to test the performance of our error correction gadget, we devise the following entanglement test. First, many perfectly encoded EPR pairs are prepared between two code-blocks of $51$ physical qubits and $17$ logical qubits each where the $X$-type, $Z$-type and logical qubits are spaced out evenly as in Section \ref{subSection:entropydensity}. This can be done by setting the input qubits associated with $X$ and $Z$ type stabilizers to $\ket{+}$ and $\ket{0}$, respectively, as before. This time, the input qubits associated with logical qubits are initialized to form an EPR pair with the corresponding input qubit of the other code-block. Now, applying the encoding circuit to each code-block separately produces encoded EPR pairs across the two code-blocks. Note that this part of the experiment is assumed to be noise-free since its purpose is to test our noisy error correction gadget.

\begin{figure}[t!]
\mbox{

\Qcircuit @C=1em @R=.7em 
{\lstick{} & \qw & \qw & \qw & \qw & \qw & \qw & \qw  & \qw & \gate{\mathcal{S}} & \multigate{1}{I(A:B)}\\
\lstick{} & \ctrl{1} \ar@{--}[]+<-.75em,.75em>;[d]+<-.75em,1em> \ar@{--}[]+<-.75em,-1em>;[l]+<-3.5em,-1em> \ar@{--}[]+<-.75em,.75em>;[r]+<6.5em,.75em> \ar@{--}[]+<8.5em,.75em>;[d]+<8.5em,-3.5em> & \targ & \qw & \qw & \qw & \qw & \cdots & & \gate{\mathcal{S}} & \ghost{I(A:B)}
    \inputgroupv{1}{2}{.8em}{1em}{\frac{\ket{\bar{00}}+\ket{\bar{11}}}{\sqrt{2}}^{\otimes k} \hspace{3em}}\\
\lstick{\ket{\bar{+}}^{\otimes k}} \ar@{--}[]+<-3.5em,.5em>;[d]+<-3.5em,-1.5em> & \targ & \qw & \meter & & \\
\lstick{\ket{\bar{0}}^{\otimes k}} \ar@{--}[]+<-3.5em,-1.5em>;[r]+<8.5em,-1.5em> & \qw & \ctrl{-2} & \gate{H} & \meter
}}
\caption{Test of our noisy error correction gadget, which is marked by the dashed box. This is the only part of the circuit that is subject to erasure errors. The ellipsis denotes that this error correction gadget is repeated many times. Each logical qubit in the first code-block is prepared in an encoded EPR pair with a corresponding logical qubit of the second code-block. This input state is denoted here as $\frac{\ket{\bar{00}}+\ket{\bar{11}}}{\sqrt{2}}^{\otimes k}$. The $\mathcal{S}$ operation represents a perfect stabilizer measurement. Finally, the quantum mutual information between the two halves of the encoded EPR pairs is calculated, which is marked by $I(A:B)$ in the diagram.}
\label{fig:midiagram}
\end{figure}
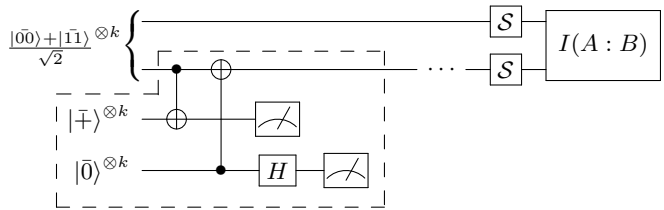

Next, the encoded states $\ket{\bar{0}}^{\otimes k}$ and $\ket{\bar{+}}^{\otimes k}$ are prepared using our noisy distillation procedure. These encoded states are then used in a Steane error correction gadget as depicted in Figure \ref{fig:steaneec}. Note this gadget is also subject to erasure errors occurring in between CNOT gates and before measurements. This entire gadget is then repeatedly applied ten times to one of the two code-blocks described previously. At the end of the last round of error correction, a perfect stabilizer measurement is performed on each code-block and the quantum mutual information between the two code-blocks is calculated. The quantum mutual information of a bipartite state $\rho_{AB}$ is defined as $I(A:B) = S(\rho_A)+S(\rho_B)-S(\rho_{AB})$ where $S(\rho)$ is the Von Neumann entropy. This entire protocol is illustrated in Figure \ref{fig:midiagram}.

The noisier the error correction gadget is, the more errors it will introduce on the EPR pairs, thus destroying the entanglement. This is captured by the quantum mutual information since it is highest when there is maximum entanglement between the two code-blocks and decreases as the entanglement is disrupted.

\begin{figure}[t!]
    \centering
    \includegraphics[width=0.95\columnwidth]{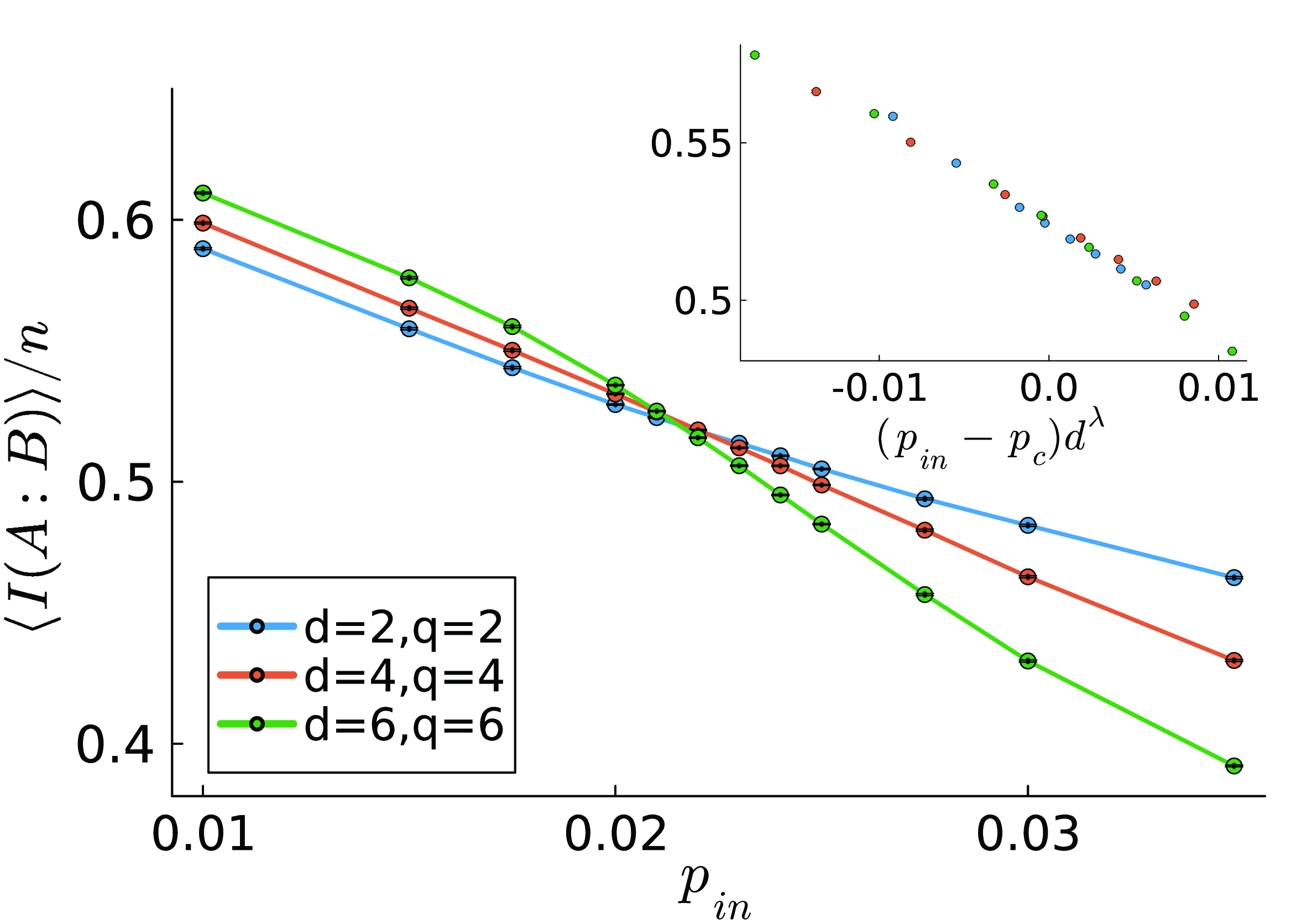}
    \caption{Mutual information density of encoded EPR pairs after noisy error correction is applied ten times. $n=51$ physical qubits are used to encode $k=17$ logical qubits which is a rate of $1/3$. $13,000$ samples were taken for points near the crossing and the standard error is displayed as error bars. The inset shows the finite-scaling collapse for $p_c = 0.0212$ and $\lambda = 0.577$. This is repeated for rates $1/4$, $1/5$, $1/10$, and $1/50$ in Figure \ref{fig:app_mi} of Appendix \ref{app:supplresults} where the corresponding estimated thresholds are $0.0216$, $0.0220$, $0.0224$, and $0.0225$, respectively.}
    \label{fig:mi}
\end{figure}

This process is repeated for randomly generated CSS circuits with varying depths. In Figure \ref{fig:mi}, depths $2$, $4$, and $6$ are used for an encoding rate of $1/3$. In each case, the number of rounds of distillation $q$ is chosen to be equal to the encoding depth $d$. For example, for depth-$2$ encoding circuits, 2 rounds of distillation are performed (one round of bit flip checks and one round of phase flip checks). Overall, 1000 different encoding circuits are used and the average and standard error of the mutual information density is plotted. In this figure, it is apparent that below a threshold error rate of about $2\%$, increasing the encoding depth and number of distillation rounds improves the performance of the error correction gadget. This suggests that below this error rate deeper encoding circuits correct more errors than they introduce. However, above this error rate it seems that deeper encoding circuits introduce more errors, even when allowing for more rounds of distillation. This is once again evidence of an error threshold for our error correction gadget. This simulation is repeated for encoding rates of $1/3$, $1/4$, $1/5$, $1/10$, and $1/50$. These results are presented in full in Figure \ref{fig:app_mi} of Appendix \ref{app:supplresults}.

\subsection{Decoding simulation using spacetime decoder}
\label{spacetime}
The last experiment that was performed is a full simulation of our scheme where erasure errors are represented as Pauli errors instead of mixed states. Performance is measured by estimating the failure rate of an optimal erasure decoder that uses the measurement outcomes and erasure error locations to determine the most likely error pattern.

In this experiment, the data block is still assumed to be noise-free and holds an arbitrary encoded state. Then, three rounds of noisy error correction  are applied followed by a perfect measurement of the code's stabilizer generators. Note that three rounds was chosen instead of ten due to the computational cost of the simulation. At each time step of the error correction gadget, erasure errors are independently applied to each qubit with probability $p$. An erasure error is simulated by applying a Pauli $X$, $Y$, or $Z$ error with equal probability and storing the location of where the error occurred. These errors then propagate throughout the circuit and end up as logical errors on the data block. The goal of the decoder is to determine what logical error occurred on the data block given the measurement outcomes and error locations. The optimal erasure decoder is implemented using ideas from the spacetime codes of Ref.~\cite{delfosse2023spacetime}, which are described below.

Recall that in the case of bit flip correction the measurement outcome of Figure \ref{fig:cnotcheck}.a is a noisy codeword of a linear code $C_1$ whose parity check matrix $H_1$ is defined by the $Z$-type stabilizers of the CSS code. More specifically, each stabilizer generator is mapped to a parity check of $H_1$ by mapping $Z$'s to $1$'s and identities to $0$'s. Furthermore, the bit flips in the noisy codeword exactly correspond to $X$ errors on the corresponding qubits. Thus the measurement outcome will be $c+e$ where $c \in C_1$ and $e$ is a bit vector representing the location of $X$ errors on the qubits being measured. To determine which $X$ errors occurred, the first step of the decoding algorithm is to calculate the error syndrome, which is given by $H_1(c+e) = H_1e$ since $c$ is a codeword. 
Phase flip correction works similarly except that the measurement outcomes are constrained by the parity check matrix defined by the $X$-type stabilizers and bit flips of these measurement outcomes correspond to $Z$ errors. 

Concatenating all measurement outcome bits into one vector $m$ and combining all of their associated parity checks into one matrix $A$ defines what is called the \textit{outcome code} of the circuit. The syndrome vector $s$ is calculated by $Am = s$. The erasure decoder is then tasked with finding an error pattern that produces the observed syndrome $s$ and is only supported on the given erasure locations. As noted in \cite{delfossezemor2020}, this is an optimal decoding strategy since all Pauli errors supported on a given set of error locations are equally likely according to the erasure error model. 

Note that in order to calculate the syndrome of an error originating at specific erasure locations it is necessary to propagate the error throughout the circuit and then check how the propagated error effects the measurement outcomes. Calculation of this error propagation is discussed further in Appendix \ref{app:spacetime}. 

To implement this decoding strategy, we use Ref.~\cite{delfosse2023spacetime}'s construction to map our circuit to a spacetime code. 
The mapping in Ref.~\cite{delfosse2023spacetime} has the advantage over the spacetime code construction of Ref.~\cite{bfhs} of being a stabilizer, rather than subsystem, code, although for general errors it does so by sacrificing some information about the syndrome that could improve the logical failure rate or threshold. 
In this mapping, each location of the circuit is represented by a physical qubit of the spacetime code. For example, a circuit with $\Delta$ layers acting on $n$-qubits has $n(\Delta + 1)$ locations (one for each qubit at each time step) and so its spacetime code will have $n(\Delta + 1)$ physical qubits. 
The key property of this spacetime code is that the syndrome of an error $E$ with respect to the spacetime code is equal to the syndrome of the same error with respect to the outcome code of the circuit. Thus, it is sufficient to find an error on the spacetime code that matches the observed syndrome and erasure locations. It is well-known that optimal erasure decoding for stabilizer codes amounts to solving a linear system of equations which can be computed efficiently. Therefore, we can use this mapping to implement an efficient erasure decoder. See Appendix \ref{app:spacetime} for more details on this erasure decoding algorithm.

\begin{figure}[t!]
    \centering
        \includegraphics[width=\linewidth]{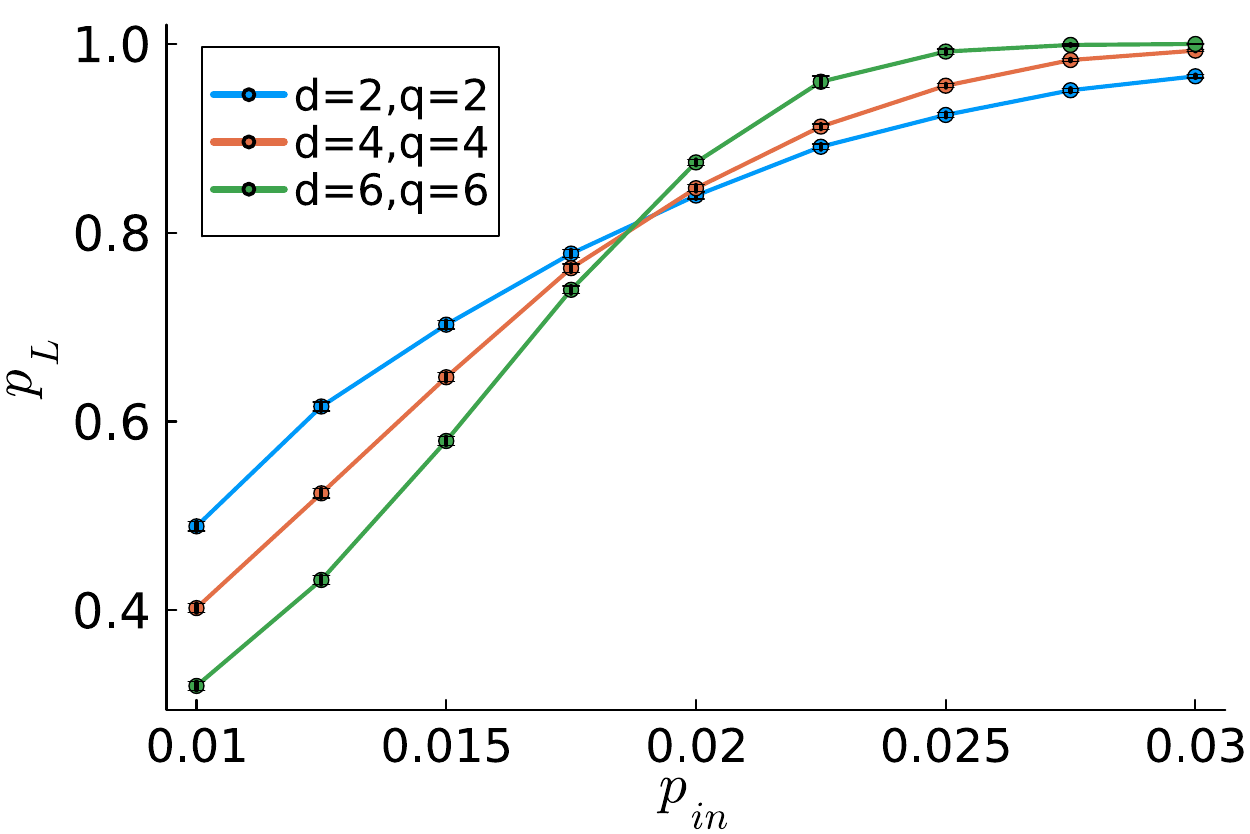}
     
    \caption{Decoding failure rate for various error rates. $51$ physical qubits are used to encode $17$ logical qubits which is a rate of $1/3$. 10,000 different CSS codes and erasure error patterns are sampled and the standard deviation of the estimated failure rate is shown as error bars.}
    \label{fig:erasure}
\end{figure}

10,000 different random CSS encoding circuits and erasure error patterns are simulated to estimate the failure rate of the error correction gadget.
Here we study the overall failure rate $p_L$ defined as the probability that any logical qubit fails. 
This is in contrast to the code capacity setting where we studied the failure rate averaged across the individual logical qubits, denoted as $\bar{p}_L$. This was needed in the code capacity setting since the open boundary conditions introduced boundary effects on the logical qubits closest to the edge. Averaging the failure rate of individual logical qubits mitigated this effect and thus helped to numerically demonstrate a threshold. In the fault-tolerant setting the encoding circuit has periodic boundary conditions and so this is no longer necessary. 

The simulation was then repeated for encoding circuits of depths $2$, $4$, and $6$ where once again the number of rounds of distillation $q$ is set to be equal to the encoding depth $d$. The results are displayed in Figure \ref{fig:erasure} where the crossing point appears close to $p=0.02$ but is a bit below the threshold estimated from the mutual information test. One potential explanation for the difference is that this decoding simulation could only be carried out for three rounds of error correction where as the mutual information simulation was performed for ten rounds.
\\

\section{Discussion}
\label{discuss}

In this paper we took steps towards the fault-tolerance of low-depth random circuit codes in noisy systems. 
By combining a state preparation protocol with Steane error correction, we showed that these random circuit codes can be used for quantum memory even when noise occurs at all stages of the error correction cycle. 
Furthermore, with limited additional overhead our protocol can be implemented with 2D local gates by leveraging transversal gates and the 1D encoding circuit. 
This geometric locality could be important for implementation on real quantum hardware; although, our protocol is also readily implemented with limited rearrangement or shuttling of qubits.
We leave a systematic analysis of geometrically local implementations with more realistic noise models for future work.

Our numerical simulations revealed threshold behavior as measured by the entropy of the encoded states, the mutual information between the quantum data and reference system, and the decoding failure rate. 
In particular, we found the erasure threshold of our scheme to be approximately $p_c = 0.02$ even for encoding rates as high as $r=1/3$. 
This result gives promising evidence that these codes can achieve a high threshold in the fault-tolerant setting even at high encoding rates. 
This desirable tradeoff between rate and threshold has previously been shown for random circuit codes in the code capacity setting, but this work takes the first step towards leveraging this property in the more realistic model of fault-tolerance.

Although our results focus on the low-depth regime, we believe our scheme can be adapted to work for deeper encoding circuits as well. Note that at high depth, it is not ideal to execute the entire encoding circuit before performing the distillation since by this point the errors may have already destroyed the state. In this case, we can instead apply rounds of distillation throughout the encoding process. This can be done by simply applying the encoding circuit in small batches of layers and running our distillation protocol after each batch until the entire encoded state is constructed. Since our experiments were restricted to the low-depth regime due to the computational cost of the numerics, we have not systematically evaluated this protocol and leave this to future work. 

Another direction for future work is to generalize our techniques to the case of circuit-level general Pauli noise in the fault-tolerant setting. We believe that similar tensor contraction methods can be used for decoding in this case.  The main issue is that the treewidth of the tensor contraction will double with each round of the distillation procedure.  To overcome this challenge, recent work has explored the use of a message-passing based approach for decoding multiple rounds of distillation for codes defined by circuits on tree geometries \cite{Sommers_2023_trees}.  We expect that similar methods can be extended to this setting with geometrically local circuit codes.

The natural remaining question is whether low-depth random circuit codes can be used for fault-tolerant quantum computation in addition to quantum memory. Since our codes are CSS codes, they have many natural fault-tolerant gadgets that we exploited here such as transversal logical CNOT gates and transversal logical measurements.  To achieve universal quantum computation, we need a method for implementing non-Clifford gates in our codes.  As of yet, there has not been a systematic investigation of logical gate constructions for low-depth random circuit codes.  However, gate-teleportation and magic-state generation protocols will have natural realizations in this setting that are compatible with the distillation procedure we used for Steane error correction here.

We also introduced practical tools for decoding low-depth random circuits. These decoders are built using tensor-network contraction and exploit the quasi-local structure of the brickwork random circuit. An important bottleneck in our decoding algorithm is the exponential scaling with circuit depth. It would be interesting to search for `sparser' codes -- low-depth circuits codes featuring stabilizers with lower average weight -- in order to mitigate this bottleneck. To do so one could consider stochastically dropping gates from the preparation circuit or by applying random projective measurements throughout the circuit \cite{li2018quantum,skinner2019measurement,chan2019unitary}. Finally, it has recently been shown by Sommers \textit{et al.} \cite{sommers2022crystalline} that a similar class of low-depth nonrandom Clifford circuits define high-performing quantum codes. In the future, we hope to further evaluate these codes using the techniques presented in this work.

\begin{acknowledgements}
We thank Grace Sommers and David Huse for discussions and collaborations on related work. We thank Daniel Gottesman for suggesting to conceptualize circuit depth as concatenation layer to help design a fault-tolerant protocol for our codes. We thank Christopher White for helpful comments on the manuscript.  This research was supported in part by NSF QLCI grant
OMA-2120757 and Grant No.\ NSF PHY-1748958. J.N. is supported by the  National Science Foundation Graduate Research Fellowship Program under Grant No. DGE  2236417.
\end{acknowledgements}

\appendix

\section{Marginal Decoding}
\label{app:marginaldecoding}
In the first two sections of the appendix, we discuss how to use our tensor contraction method to decode finite-rate random circuit codes. One strategy for decoding finite-rate codes is to iterate through each logical qubit and assign it to the logical class with the largest marginal probability. Before formally defining this marginal probability, we introduce some notation. The logical generators are comprised of the operators $L^X_j$ and $L^Z_j$ which together generate the full logical group $L$ and represent the logical $X$ and $Z$ operators, respectively, on logical qubit $j$. Now, we can define the set of logical generators on all qubits besides the $j$th logical qubit as 
\begin{equation}
    G_{-j} = \{L^X_{j'} | j' \neq j \} \cup \{L^Z_{j'} | j' \neq j \}.
\end{equation}
We let $L_{-j} = \langle G_{-j} \rangle$ be the set of logical operators that is generated by $G_{-j}$. Finally, we let $L^{\sigma}_j$ be the set of single-qubit logical classes of the $j$th logical qubit, i.e. the set of logical operators generated by $L^X_j$ and $L^Z_j$. With this notation in mind, the marginal probability of the logical qubit $j$ being $L^{\sigma}_j$ is
\begin{equation}
    \label{marg}
    P_j(\sigma) = \sum_{\substack{S_a \in S \\ L_{-j,\ell} \in L_{-j}}}P(S_a L_{-j,\ell} L^{\sigma}_j C_s)
\end{equation}

These marginal probabilities can be computed by adding all logical generators in $G_{-j}$ to the Hamiltonian and calculating the partition function. 
For Pauli error $E$, stabilizer generators $S_1$,...,$S_{n-k}$, and logical $j$, the classical spin Hamiltonian is redefined as
\begin{align}
    H_{E,j}(\vec{s}) =
    \sum_{\sigma} & J_\sigma \llbracket E, \sigma \rrbracket  \prod_{ \llbracket \sigma, S_i \rrbracket = -1} s_i \prod_{\llbracket \sigma, G_{-j,i'} \rrbracket = -1} g_{i'}, 
\end{align}
where $g_{i'} = \pm 1$ are a new set of Ising spins, one for each generator $G_{-j,i'}$ of $G_{-j}$, and $\sigma$ is any weight-one $n$-qubit Pauli operator. The model now contains $n-k+2k-2 = n+k-2$ spins since there is one for each stabilizer and logical generator excluding the two logical generators on $j$.

The partition function is now a sum over not only all of the stabilizer elements but all of the logical operators in $L_{-j}$ multiplied by $E$. Thus we now have,
\begin{align}
    Z_{E,j} &= \sum_{\vec{s}} e^{-\beta H_{E,j}(\vec{s})} \propto \sum_{\substack{S_a \in S \\ L_{-j,\ell} \in L_{-j}}}P(S_a L_{-j,\ell} E). 
\end{align}
By setting $E = L^{\sigma}_j C_s$ and combining with Eq. \ref{marg}, we see that $Z_{E,j}$ is proportional to $P_j(\sigma)$. Finally, choosing $\sigma \in \{I, X, Y, Z\}$ such that  $Z_{E,j}$ is maximized will give the logical class that optimizes the marginal probability.

\section{Minimum Weight Decoding}
\label{app:minweightdecoding}

Here we describe minimum weight decoding, a sub-optimal alternative to maximum-likelihood decoding that is typically more efficient in practice.
We discuss this class of decoders using the statistical mechanics framework introduced in Section \ref{decoding}. To perform minimum weight decoding, the temperature of the system is set to zero instead of according to the Nishimori conditions. Since the partition function is not well-defined in this case, the free energy is used instead:
\begin{equation}
F_E(\beta) = -\frac{1}{\beta} \log Z_E.
\end{equation}
Notice that this is a monotonic function and so minimizing the free energy and maximizing the partition function are equivalent. The difference between maximum likelihood and minimum weight decoding is therefore entirely captured by the shift in temperature. Notice that the free energy at zero temperature reduces to the ground state energy:
\begin{align}
\lim_{\beta \to \infty} F_E(\beta) &= -\lim_{\beta \to \infty}\frac{1}{\beta} \log Z_E \\ 
&= -\lim_{\beta \to \infty}\frac{1}{\beta} \log \sum_{\vec{s}} e^{-\beta H_E(\vec{s})} \\ 
&= \min_{\vec{s}} H_{E}(\vec{s}).
\label{groundstate}
\end{align}

This ground state is exactly the minimum weight Pauli operator reachable by multiplying $E$ by the generators associated with the spin variables. To see this, first notice that there are three Hamiltonian terms corresponding to each qubit $i$ of $E$ (one for each single-weight operator $X_i, Y_i$, and $Z_i$). For simplicity, let us first consider the case where $\vec{s} = \vec{1}$. When $E_i = I$, these three Hamiltonian terms have a total energy bonus of $-3$ since in this case $E_i$ commutes with all three single-weight Pauli operators. On the other hand, these Hamiltonian terms have a total energy penalty of $+1$ when $E_i$ is nontrivial since it now anticommutes with two out of the three $X_i$, $Y_i$, $Z_i$ operators. Therefore, the energy of the Hamiltonian is completely determined by how many qubits of $E$ are nontrivial. Now recall that flipping a spin from $1$ to $-1$ is equivalent to multiplying the error $E$ by the corresponding generator. Thus, the spin configuration with the lowest energy corresponds to the subset of generators that produces the minimum weight operator when multiplied together with $E$. If all stabilizer and logical generators are included as spins $s_1,\ldots,s_{n-k},\ldots,s_{n+k}$ in the Hamiltonian as shown below
\begin{equation*}
\begin{split}
    H_{E}(\vec{s}) = \sum_{\sigma} J_\sigma \llbracket E, \sigma \rrbracket & \prod_{\llbracket \sigma, S_i \rrbracket = -1} s_i 
     \prod_{\llbracket \sigma, L^X_{i'} \rrbracket = -1} s_{n-k+i'} \\
     &\times\prod_{\llbracket \sigma, L^Z_{i''} \rrbracket = -1} s_{n +i''},
    \end{split}
\end{equation*}
then the equivalence between finding the ground state and minimum weight decoding can be made explicit as

\begin{equation}
	\min_{\vec{s}} H_{E}(\vec{s}) = \min_{\substack{S_a \in S \\ L_{\ell } \in L}} \text{wt}(S_a L_\ell E) . 
\end{equation}
where $L$ is the set of all logical operations. When $E = C_s$, it is apparent that this is the minimum weight error that is consistent with the syndrome $s$.

\section{Efficient Tensor Network Contraction}
\label{app:tensor}
\subsection{Calculating the partition function}
It is well-known that the partition function of a classical spin model can be written as a tensor network contraction. Although it is \#P-hard to contract the resulting tensor network in general, the hope is that the low-depth structure of the encoding circuit can be leveraged to perform this efficiently. To rewrite the partition function as a tensor network contraction, we follow the notation of Ref.~\cite{chubb2021statistical} and introduce the tensor variables $\alpha_{iu}$ where $i$ and $u$ are the spin and Hamiltonian term labels, respectively. To enforce that each variable representing the same spin takes the same value, delta tensors denoted $\delta_i$ are included that take as input all $\alpha$ variables with spin label $i$ and output one only when they each have the same value; otherwise, zero is outputted. With this change of variables the partition function is expressed as 
\begin{equation}
\label{eq:partitiontensor}
    Z_E = \sum_{\alpha_{iu}} \prod_{i'} \delta_{i'}(\{\alpha_{iu} | i = i'\}) \prod_{u'} e^{-\beta h_{u'}(\{\alpha_{iu} | u = u'\})}
\end{equation}
where $h_u$ denotes each Hamiltonian term and is a function of all variables of $\alpha$ that are labeled by $u$.

\begin{figure}
    \centering
    \includegraphics[width=\columnwidth]{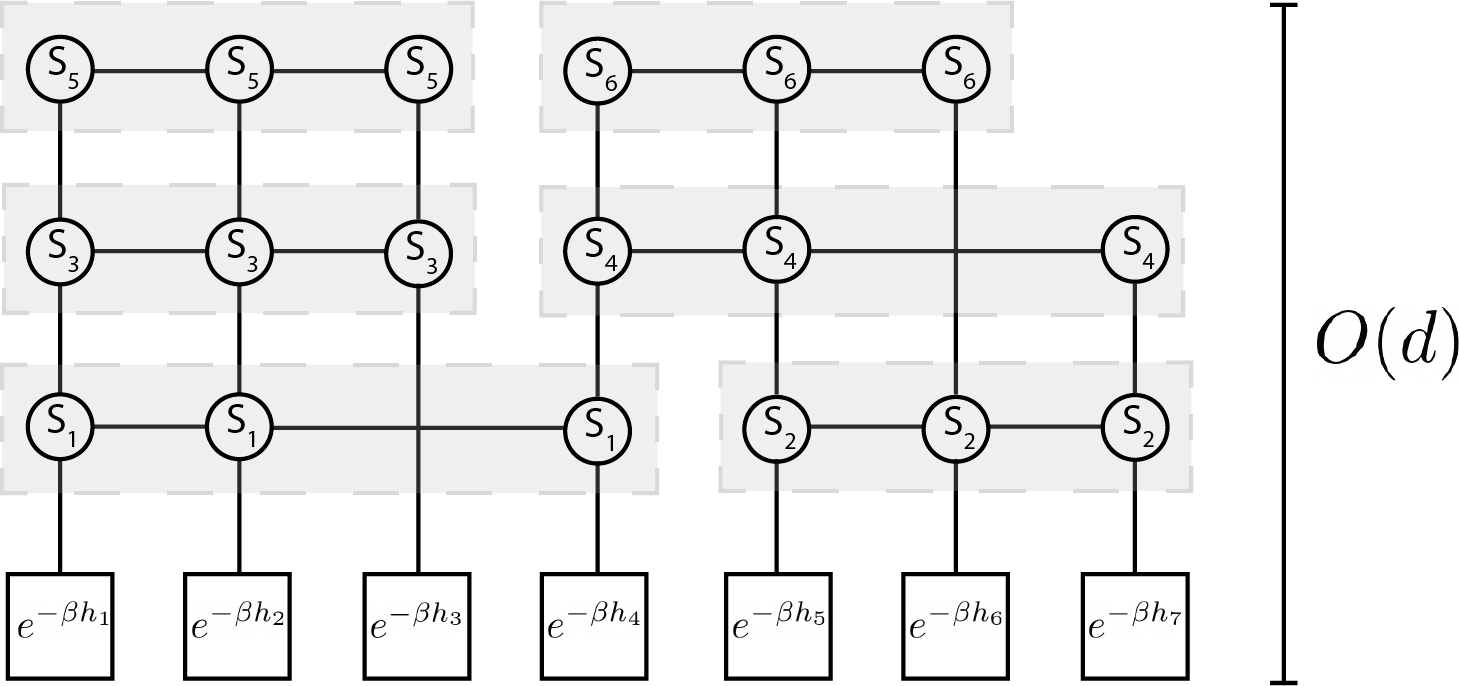}
    \caption{Diagram of 2D tensor network. Each horizontal chain of $s_i$'s represents one spin in the stat mech model. For any tensor in this chain, it is enforced that both horizontal edges have the same value otherwise the tensor outputs zero. This ensures that the chain acts as one spin. The vertical edges play the role of multiplying the spins together. A given tensor will output one if its spin value, defined by the horizontal edge, multiplied by the previous spin value, defined by the above vertical edge, is equal to the outgoing spin value, defined by the below vertical edge. The vertical edge incident to the Hamiltonian terms, denoted by squares, now represents the product of all of the spin values. The square tensor outputs $e^{-\beta J \llbracket E, \sigma \rrbracket \prod_i s_i}$ given $\prod_i s_i$ from the incoming edge.}
    \label{fig:2dtensor}
\end{figure}

This tensor network can be visualized as a bipartite graph with the delta tensors on one side and the Hamiltonian terms on the other. 
Using a similar method as Ref.~\cite{darmawan2022low}, we make the contraction more efficient by transforming the network into a 2D lattice with height $O(d)$ where $d$ is the depth of the encoding circuit as shown in Figure \ref{fig:2dtensor}. This 2D lattice can now be contracted efficiently by traversing it column by column. With this ordering, the runtime of the contraction is exponential only in the height, which is still polynomial when the depth is logarithmic in the number of qubits. To do this transformation to a 2D lattice, each $\delta_i$ tensor is broken up into a chain of smaller tensors that each correspond to a different incident Hamiltonian term as illustrated in Fig. \ref{fig:2dtensor}. It is enforced that each of the new $\delta$ tensors in the chain only outputs one if the neighboring $\alpha$ variables are all the same, and so the overall function of the delta tensors is unchanged. The next modification is that all delta tensors that are incident to the same $e^{-\beta h_u}$ tensor are stacked on top of each other in a vertical chain. The vertical tensor legs represent the product of the spin variables of a given Hamiltonian term and this value is updated according to $\alpha_{iu}$, which are passed by the horizontal tensor legs. Finally, once the bottom is reached, the product of spin variables is passed to the $e^{-\beta h_u}$ tensor. In this way, the tensor network still represents the original partition function calculation, but now takes the form of a 2D lattice. 

The height of the network depends on the total number of delta tensor chains that must be stacked on top of each other. Since each chain of delta tensors represents a spin, this is at most the total number of spins in a single Hamiltonian term. To see that this is bounded by $O(d)$ it is first crucial to note that the reverse light-cone of the encoding circuit is $2d$ since each layer of gates causes the light-cone to grow by at most one qubit on either side. Therefore, at most $O(d)$ generators can be incident to a given qubit and so each Hamiltonian term can have at most $O(d)$ spins. This guarantees that the height will always be bounded by $O(d)$ as well.
\subsection{Finding the ground state using tropical tensor networks}
Recall from Eq. \ref{groundstate} that in the zero temperature limit the free energy reduces to the ground state energy. In order to carry out this calculation we use numerical techniques presented in Ref.~\cite{liu2021tropical}, which make use of two important observations:
\begin{equation}
    \label{tropadd}
    -\lim_{\beta \to \infty} \frac{1}{\beta} \log (e^{-\beta x} + e^{-\beta y}) = \min(x, y)
\end{equation} and
\begin{equation}
    \label{tropmult}
    -\lim_{\beta \to \infty} \frac{1}{\beta} \log (e^{-\beta x} e^{-\beta y}) = x+y.
\end{equation} 

Combining Eq. \ref{groundstate} and Eq. \ref{eq:partitiontensor} we can write the ground state energy as 
\begin{equation*}
    \label{gsetensor}
    \begin{split}
    \min_{\vec{s}} H_{E}(\vec{s}) = -\lim_{\beta \to \infty} \frac{1}{\beta} &\log \Big[\sum_{\alpha_{iu}} \prod_{i'} \delta_{i'}(\{\alpha_{iu} | i = i'\}) \\
    &\times \prod_{u'} e^{-\beta h_{u'}(\{\alpha_{iu} | u = u'\})}\Big]
    \end{split}
\end{equation*}
Inspired by Eq. \ref{tropadd} and \ref{tropmult}, we introduce a new algebra in which multiplication and addition are defined as follows:
\begin{equation}
    a \odot b = a+b
\end{equation}
\begin{equation}
    a \oplus b = \min(a, b).
\end{equation}
This is known as the tropical algebra. In this algebra, $0$ acts as the unit element since $e^{-\beta (0)} = 1$ and $\infty$ acts as the zero element since $e^{-\beta(\infty)} = 0$.  
Using these tropical definitions for multiplication and addition, we can now define the ground state energy as a tensor contraction in the tropical algebra:

\begin{align}
    &\min_{\vec{s}} H_{E}(\vec{s}) = \\
    &\bigoplus_{\alpha_{iu}} \bigodot_{i'} \delta_{i'}(\{\alpha_{iu} | i = i'\}) \bigodot_{u'} h_{u'}(\{\alpha_{iu} | u = u'\})). \nonumber
\end{align}
This can be contracted efficiently using the same transformation to a 2D lattice as described previously. 

However, this only calculates the ground state energy rather than the actual ground state. To get the ground-state, two different methods can be used. First, a small perturbative field term can be added to the Hamiltonian for each spin. Now, differentiating the tensor contraction with respect to the spin variables will give the ground state configuration. This differentiation step will incur an $O(n)$ overhead in runtime. Our method instead trades off the runtime overhead for an $O(n)$ memory overhead by storing the ground state configuration as the tensor network is contracted.

\section{Spacetime Decoder}
\label{app:spacetime}
In this section, the spacetime erasure decoder is presented in detail. First, we introduce notation from Ref.~\cite{delfosse2023spacetime} to formalize the algorithm. Given a Clifford circuit $\mathcal{C}$ with $\Delta$ layers of gates acting on $n$ qubits, a \textit{fault operator }$F$ is defined as an $n(\Delta+1)$-qubit Pauli operator where each qubit of $F$ corresponds to a qubit at a specific time step in the circuit. These fault operators represent error patterns occurring in the circuit. Let $\ell \in \{0, 1, \dots, \Delta\}$ index each level of the circuit where $\ell=0$ refers to the level with no gates applied. Then the time step in between level $\ell$ and $\ell+1$ is denoted as $\ell+0.5$ and the $n$-qubit Pauli error occurring on the qubits at time step $\ell+0.5$ is denoted $F_{\ell+0.5}$. We also denote each layer of the circuit as $U_\ell$ and the product of layers in between level $i$ and $j$ as $U_{i,j} = U_j U_{j-1}\dots U_{i+1}$. Finally, the propagation of a fault operator $F$ through the circuit can be defined as the \textit{cumulant} of $F$, denoted as $\overrightarrow{F}$. The explicit formula for the cumulant is reproduced from Proposition 2 of Ref.~\cite{delfosse2023spacetime} below.
\begin{align}
    \overrightarrow{F}_{\ell+0.5} = \prod^\ell_{i=0} U_{i,\ell}F_{i+0.5}U_{i,\ell}^\dagger
\end{align}
 Essentially all faults from previous levels are propagated to the current level and multiplied together. The back-cumulant can similarly be defined by propagating the fault backwards.
 \begin{align}
    \overleftarrow{F}_{\ell+0.5} = \prod^{\Delta}_{j=\ell} U_{\ell,j}^\dagger F_{j+0.5} U_{\ell,j}
\end{align}

Next, we describe how the stabilizer generators of the spacetime code are defined. Recall that the measurement outcomes of the circuit are constrained by the parity checks of the circuit's outcome code. In particular, a nonzero inner product between a parity check and the measurement outcome vector indicates an error has occurred. For example, for standard basis measurements, an $X$ error at the time of measurement will flip the measurement outcome and can be detected by the outcome code parity checks. However, a $Z$ error does not change the measurement result and, thus, goes undetected. 

As discussed in Section \ref{spacetime}, the parity checks are defined by the CSS code's stabilizer generators. For the bit flip correction subcircuit, a parity check is added to the outcome code for each $Z$-type stabilizer of the measured code-block. In particular, each single-qubit $Z$ operator of the generator is mapped to a $1$ of the parity check at the corresponding measurement bit. Similarly, $I$'s of the stabilizer generator are mapped to $0$'s of the parity check. The parity checks for the Hadamard basis measurements in the phase flip correction subcircuits are defined analogously using the $X$-type stabilizers. 

A key insight is that a parity check has nonzero inner product with the measurement outcome vector if and only if the corresponding stabilizer generator anticommutes with the error.
To see this, first note that the measurement outcome vector can be written as $c+e$ where $c\in C_1$ for the code $C_1$ defined by the parity check matrix of the outcome code denoted here as $H_1$. The measurement bit flips caused by the error are represented as $e$. For standard basis measurements in the bit flip correction subcircuit, $e$ exactly corresponds to $X$ errors on the corresponding qubits and $H_1$ corresponds to the binary representation of the $Z$-type stabilizer generators. Similarly, for Hadamard basis measurements in the phase flip correction circuit, $e$ corresponds to $Z$ errors and $H_1$ corresponds to the binary representation of the $X$-type stabilizer generators. Finally, note that the syndrome with respect to the outcome code is $H_1(c+e) = H_1e$, which is equivalent to the commutator of the Pauli error with the stabilizer generators associated with $H_1$.

Therefore, the outcome code can be mapped to a stabilizer code called the spacetime code by including the corresponding stabilizer generator of each parity check as a generator. More specifically, each parity check is mapped to a stabilizer generator of the spacetime code by setting each location of the circuit other than the given measurement locations to identity and setting these measurement locations to be the corresponding stabilizer generator of the original code.

For the spacetime stabilizer generators $\{S_i\}$ defined in this way, we have that for a given error pattern $E$ the syndrome with respect to the outcome code is exactly $s_i = \llbracket S_i,\overrightarrow{E}\rrbracket$. This is not quite the desired mapping since the syndrome is only correct after propagating $E$ through the circuit. This can be fixed by applying Proposition 3 of Ref.~\cite{delfosse2023spacetime}, which shows that $\llbracket F,\overrightarrow{G}\rrbracket = \llbracket \overleftarrow{F},G\rrbracket$. Thus we can replace our stabilizer generators $\{S_i\}$ with $\{\overleftarrow{S_i}\}$ so that $s_i = \llbracket \overleftarrow{S_i},E\rrbracket$. Now, erasure decoding can be applied to this spacetime code at the given erasure error locations. This will return a Pauli error that is supported on the erasure locations and produces the given syndrome $s$ with respect to the outcome code when propagated throughout the circuit.

\bibliography{References.bib}

\begin{thebibliography}{58}%
\makeatletter
\providecommand \@ifxundefined [1]{%
 \@ifx{#1\undefined}
}%
\providecommand \@ifnum [1]{%
 \ifnum #1\expandafter \@firstoftwo
 \else \expandafter \@secondoftwo
 \fi
}%
\providecommand \@ifx [1]{%
 \ifx #1\expandafter \@firstoftwo
 \else \expandafter \@secondoftwo
 \fi
}%
\providecommand \natexlab [1]{#1}%
\providecommand \enquote  [1]{``#1''}%
\providecommand \bibnamefont  [1]{#1}%
\providecommand \bibfnamefont [1]{#1}%
\providecommand \citenamefont [1]{#1}%
\providecommand \href@noop [0]{\@secondoftwo}%
\providecommand \href [0]{\begingroup \@sanitize@url \@href}%
\providecommand \@href[1]{\@@startlink{#1}\@@href}%
\providecommand \@@href[1]{\endgroup#1\@@endlink}%
\providecommand \@sanitize@url [0]{\catcode `\\12\catcode `\$12\catcode
  `\&12\catcode `\#12\catcode `\^12\catcode `\_12\catcode `\%12\relax}%
\providecommand \@@startlink[1]{}%
\providecommand \@@endlink[0]{}%
\providecommand \url  [0]{\begingroup\@sanitize@url \@url }%
\providecommand \@url [1]{\endgroup\@href {#1}{\urlprefix }}%
\providecommand \urlprefix  [0]{URL }%
\providecommand \Eprint [0]{\href }%
\providecommand \doibase [0]{http://dx.doi.org/}%
\providecommand \selectlanguage [0]{\@gobble}%
\providecommand \bibinfo  [0]{\@secondoftwo}%
\providecommand \bibfield  [0]{\@secondoftwo}%
\providecommand \translation [1]{[#1]}%
\providecommand \BibitemOpen [0]{}%
\providecommand \bibitemStop [0]{}%
\providecommand \bibitemNoStop [0]{.\EOS\space}%
\providecommand \EOS [0]{\spacefactor3000\relax}%
\providecommand \BibitemShut  [1]{\csname bibitem#1\endcsname}%
\let\auto@bib@innerbib\@empty
\bibitem [{\citenamefont {Darmawan}\ \emph {et~al.}(2022)\citenamefont
  {Darmawan}, \citenamefont {Nakata}, \citenamefont {Tamiya},\ and\
  \citenamefont {Yamasaki}}]{darmawan2022low}%
  \BibitemOpen
  \bibfield  {author} {\bibinfo {author} {\bibfnamefont {A.~S.}\ \bibnamefont
  {Darmawan}}, \bibinfo {author} {\bibfnamefont {Y.}~\bibnamefont {Nakata}},
  \bibinfo {author} {\bibfnamefont {S.}~\bibnamefont {Tamiya}}, \ and\ \bibinfo
  {author} {\bibfnamefont {H.}~\bibnamefont {Yamasaki}},\ }\href@noop {}
  {\bibfield  {journal} {\bibinfo  {journal} {arXiv preprint arXiv:2212.05071}\
  } (\bibinfo {year} {2022})}\BibitemShut {NoStop}%
\bibitem [{\citenamefont {Shannon}(1948)}]{shannon1948mathematical}%
  \BibitemOpen
  \bibfield  {author} {\bibinfo {author} {\bibfnamefont {C.~E.}\ \bibnamefont
  {Shannon}},\ }\href@noop {} {\bibfield  {journal} {\bibinfo  {journal} {The
  Bell system technical journal}\ }\textbf {\bibinfo {volume} {27}},\ \bibinfo
  {pages} {379} (\bibinfo {year} {1948})}\BibitemShut {NoStop}%
\bibitem [{\citenamefont {Gallager}(1962)}]{gallager1962low}%
  \BibitemOpen
  \bibfield  {author} {\bibinfo {author} {\bibfnamefont {R.}~\bibnamefont
  {Gallager}},\ }\href@noop {} {\bibfield  {journal} {\bibinfo  {journal} {IRE
  Transactions on information theory}\ }\textbf {\bibinfo {volume} {8}},\
  \bibinfo {pages} {21} (\bibinfo {year} {1962})}\BibitemShut {NoStop}%
\bibitem [{\citenamefont {MacKay}\ and\ \citenamefont
  {Neal}(1997)}]{mackay1997near}%
  \BibitemOpen
  \bibfield  {author} {\bibinfo {author} {\bibfnamefont {D.~J.}\ \bibnamefont
  {MacKay}}\ and\ \bibinfo {author} {\bibfnamefont {R.~M.}\ \bibnamefont
  {Neal}},\ }\href@noop {} {\bibfield  {journal} {\bibinfo  {journal}
  {Electronics letters}\ }\textbf {\bibinfo {volume} {33}},\ \bibinfo {pages}
  {457} (\bibinfo {year} {1997})}\BibitemShut {NoStop}%
\bibitem [{\citenamefont {Masera}\ \emph {et~al.}(2007)\citenamefont {Masera},
  \citenamefont {Quaglio},\ and\ \citenamefont
  {Vacca}}]{masera2007implementation}%
  \BibitemOpen
  \bibfield  {author} {\bibinfo {author} {\bibfnamefont {G.}~\bibnamefont
  {Masera}}, \bibinfo {author} {\bibfnamefont {F.}~\bibnamefont {Quaglio}}, \
  and\ \bibinfo {author} {\bibfnamefont {F.}~\bibnamefont {Vacca}},\ }\href
  {\doibase 10.1109/TCSII.2007.894409} {\bibfield  {journal} {\bibinfo
  {journal} {IEEE Transactions on Circuits and Systems II: Express Briefs}\
  }\textbf {\bibinfo {volume} {54}},\ \bibinfo {pages} {542} (\bibinfo {year}
  {2007})}\BibitemShut {NoStop}%
\bibitem [{\citenamefont {Gottesman}(2013)}]{gottesman2013fault}%
  \BibitemOpen
  \bibfield  {author} {\bibinfo {author} {\bibfnamefont {D.}~\bibnamefont
  {Gottesman}},\ }\href@noop {} {\bibfield  {journal} {\bibinfo  {journal}
  {arXiv preprint arXiv:1310.2984}\ } (\bibinfo {year} {2013})}\BibitemShut
  {NoStop}%
\bibitem [{\citenamefont {Kovalev}\ and\ \citenamefont
  {Pryadko}(2013)}]{kovalev2013fault}%
  \BibitemOpen
  \bibfield  {author} {\bibinfo {author} {\bibfnamefont {A.~A.}\ \bibnamefont
  {Kovalev}}\ and\ \bibinfo {author} {\bibfnamefont {L.~P.}\ \bibnamefont
  {Pryadko}},\ }\href {\doibase 10.1103/PhysRevA.87.020304} {\bibfield
  {journal} {\bibinfo  {journal} {Phys. Rev. A}\ }\textbf {\bibinfo {volume}
  {87}},\ \bibinfo {pages} {020304} (\bibinfo {year} {2013})}\BibitemShut
  {NoStop}%
\bibitem [{\citenamefont {Brown}\ and\ \citenamefont
  {Fawzi}(2013)}]{brown2013short}%
  \BibitemOpen
  \bibfield  {author} {\bibinfo {author} {\bibfnamefont {W.}~\bibnamefont
  {Brown}}\ and\ \bibinfo {author} {\bibfnamefont {O.}~\bibnamefont {Fawzi}},\
  }in\ \href {\doibase 10.1109/ISIT.2013.6620245} {\emph {\bibinfo {booktitle}
  {2013 IEEE International Symposium on Information Theory}}}\ (\bibinfo {year}
  {2013})\ pp.\ \bibinfo {pages} {346--350}\BibitemShut {NoStop}%
\bibitem [{\citenamefont {Brown}\ and\ \citenamefont
  {Fawzi}(2015)}]{brown2015decoupling}%
  \BibitemOpen
  \bibfield  {author} {\bibinfo {author} {\bibfnamefont {W.}~\bibnamefont
  {Brown}}\ and\ \bibinfo {author} {\bibfnamefont {O.}~\bibnamefont {Fawzi}},\
  }\href@noop {} {\bibfield  {journal} {\bibinfo  {journal} {Communications in
  mathematical physics}\ }\textbf {\bibinfo {volume} {340}},\ \bibinfo {pages}
  {867} (\bibinfo {year} {2015})}\BibitemShut {NoStop}%
\bibitem [{\citenamefont {Fawzi}\ \emph {et~al.}(2020)\citenamefont {Fawzi},
  \citenamefont {Grospellier},\ and\ \citenamefont
  {Leverrier}}]{fawzi2020constant}%
  \BibitemOpen
  \bibfield  {author} {\bibinfo {author} {\bibfnamefont {O.}~\bibnamefont
  {Fawzi}}, \bibinfo {author} {\bibfnamefont {A.}~\bibnamefont {Grospellier}},
  \ and\ \bibinfo {author} {\bibfnamefont {A.}~\bibnamefont {Leverrier}},\
  }\href {\doibase 10.1145/3434163} {\bibfield  {journal} {\bibinfo  {journal}
  {Commun. ACM}\ }\textbf {\bibinfo {volume} {64}},\ \bibinfo {pages}
  {106–114} (\bibinfo {year} {2020})}\BibitemShut {NoStop}%
\bibitem [{\citenamefont {Gullans}\ \emph {et~al.}(2021)\citenamefont
  {Gullans}, \citenamefont {Krastanov}, \citenamefont {Huse}, \citenamefont
  {Jiang},\ and\ \citenamefont {Flammia}}]{gullans2021quantum}%
  \BibitemOpen
  \bibfield  {author} {\bibinfo {author} {\bibfnamefont {M.~J.}\ \bibnamefont
  {Gullans}}, \bibinfo {author} {\bibfnamefont {S.}~\bibnamefont {Krastanov}},
  \bibinfo {author} {\bibfnamefont {D.~A.}\ \bibnamefont {Huse}}, \bibinfo
  {author} {\bibfnamefont {L.}~\bibnamefont {Jiang}}, \ and\ \bibinfo {author}
  {\bibfnamefont {S.~T.}\ \bibnamefont {Flammia}},\ }\href {\doibase
  10.1103/PhysRevX.11.031066} {\bibfield  {journal} {\bibinfo  {journal} {Phys.
  Rev. X}\ }\textbf {\bibinfo {volume} {11}},\ \bibinfo {pages} {031066}
  (\bibinfo {year} {2021})}\BibitemShut {NoStop}%
\bibitem [{\citenamefont {Leverrier}\ and\ \citenamefont
  {Z{\'e}mor}(2022)}]{leverrier2022quantum}%
  \BibitemOpen
  \bibfield  {author} {\bibinfo {author} {\bibfnamefont {A.}~\bibnamefont
  {Leverrier}}\ and\ \bibinfo {author} {\bibfnamefont {G.}~\bibnamefont
  {Z{\'e}mor}},\ }in\ \href@noop {} {\emph {\bibinfo {booktitle} {2022 IEEE
  63rd Annual Symposium on Foundations of Computer Science (FOCS)}}}\ (\bibinfo
  {organization} {IEEE},\ \bibinfo {year} {2022})\ pp.\ \bibinfo {pages}
  {872--883}\BibitemShut {NoStop}%
\bibitem [{\citenamefont {Hayden}\ \emph {et~al.}(2008)\citenamefont {Hayden},
  \citenamefont {Horodecki}, \citenamefont {Winter},\ and\ \citenamefont
  {Yard}}]{hayden2008decoupling}%
  \BibitemOpen
  \bibfield  {author} {\bibinfo {author} {\bibfnamefont {P.}~\bibnamefont
  {Hayden}}, \bibinfo {author} {\bibfnamefont {M.}~\bibnamefont {Horodecki}},
  \bibinfo {author} {\bibfnamefont {A.}~\bibnamefont {Winter}}, \ and\ \bibinfo
  {author} {\bibfnamefont {J.}~\bibnamefont {Yard}},\ }\href {\doibase
  10.1142/S1230161208000043} {\bibfield  {journal} {\bibinfo  {journal} {Open
  Systems \& Information Dynamics}\ }\textbf {\bibinfo {volume} {15}},\
  \bibinfo {pages} {7} (\bibinfo {year} {2008})},\ \Eprint
  {http://arxiv.org/abs/https://doi.org/10.1142/S1230161208000043}
  {https://doi.org/10.1142/S1230161208000043} \BibitemShut {NoStop}%
\bibitem [{\citenamefont {Kitaev}(1997)}]{Kitaev_1997}%
  \BibitemOpen
  \bibfield  {author} {\bibinfo {author} {\bibfnamefont {A.~Y.}\ \bibnamefont
  {Kitaev}},\ }\href {\doibase 10.1070/RM1997v052n06ABEH002155} {\bibfield
  {journal} {\bibinfo  {journal} {Russian Mathematical Surveys}\ }\textbf
  {\bibinfo {volume} {52}},\ \bibinfo {pages} {1191} (\bibinfo {year}
  {1997})}\BibitemShut {NoStop}%
\bibitem [{\citenamefont {Tillich}\ and\ \citenamefont
  {Zemor}(2014)}]{Tillich_2014}%
  \BibitemOpen
  \bibfield  {author} {\bibinfo {author} {\bibfnamefont {J.-P.}\ \bibnamefont
  {Tillich}}\ and\ \bibinfo {author} {\bibfnamefont {G.}~\bibnamefont
  {Zemor}},\ }\href {\doibase 10.1109/tit.2013.2292061} {\bibfield  {journal}
  {\bibinfo  {journal} {{IEEE} Transactions on Information Theory}\ }\textbf
  {\bibinfo {volume} {60}},\ \bibinfo {pages} {1193} (\bibinfo {year}
  {2014})}\BibitemShut {NoStop}%
\bibitem [{\citenamefont {Kovalev}\ and\ \citenamefont
  {Pryadko}(2012)}]{Kovalev_2012}%
  \BibitemOpen
  \bibfield  {author} {\bibinfo {author} {\bibfnamefont {A.~A.}\ \bibnamefont
  {Kovalev}}\ and\ \bibinfo {author} {\bibfnamefont {L.~P.}\ \bibnamefont
  {Pryadko}},\ }in\ \href {\doibase 10.1109/isit.2012.6284206} {\emph {\bibinfo
  {booktitle} {2012 {IEEE} International Symposium on Information Theory
  Proceedings}}}\ (\bibinfo  {publisher} {{IEEE}},\ \bibinfo {year}
  {2012})\BibitemShut {NoStop}%
\bibitem [{\citenamefont {Delfosse}\ \emph {et~al.}(2016)\citenamefont
  {Delfosse}, \citenamefont {Iyer},\ and\ \citenamefont {Poulin}}]{Delfosse16}%
  \BibitemOpen
  \bibfield  {author} {\bibinfo {author} {\bibfnamefont {N.}~\bibnamefont
  {Delfosse}}, \bibinfo {author} {\bibfnamefont {P.}~\bibnamefont {Iyer}}, \
  and\ \bibinfo {author} {\bibfnamefont {D.}~\bibnamefont {Poulin}},\
  }\href@noop {} {\  (\bibinfo {year} {2016})},\ \Eprint
  {http://arxiv.org/abs/arXiv:1611.04256} {arXiv:1611.04256} \BibitemShut
  {NoStop}%
\bibitem [{\citenamefont {Delfosse}\ and\ \citenamefont
  {Z\'emor}(2020{\natexlab{a}})}]{Delfosse17}%
  \BibitemOpen
  \bibfield  {author} {\bibinfo {author} {\bibfnamefont {N.}~\bibnamefont
  {Delfosse}}\ and\ \bibinfo {author} {\bibfnamefont {G.}~\bibnamefont
  {Z\'emor}},\ }\href {\doibase 10.1103/PhysRevResearch.2.033042} {\bibfield
  {journal} {\bibinfo  {journal} {Phys. Rev. Research}\ }\textbf {\bibinfo
  {volume} {2}},\ \bibinfo {pages} {033042} (\bibinfo {year}
  {2020}{\natexlab{a}})},\ \Eprint {http://arxiv.org/abs/arXiv:1703.01517}
  {arXiv:1703.01517} \BibitemShut {NoStop}%
\bibitem [{\citenamefont {Wu}\ \emph {et~al.}(2022)\citenamefont {Wu},
  \citenamefont {Kolkowitz}, \citenamefont {Puri},\ and\ \citenamefont
  {Thompson}}]{Wu_2022}%
  \BibitemOpen
  \bibfield  {author} {\bibinfo {author} {\bibfnamefont {Y.}~\bibnamefont
  {Wu}}, \bibinfo {author} {\bibfnamefont {S.}~\bibnamefont {Kolkowitz}},
  \bibinfo {author} {\bibfnamefont {S.}~\bibnamefont {Puri}}, \ and\ \bibinfo
  {author} {\bibfnamefont {J.~D.}\ \bibnamefont {Thompson}},\ }\href {\doibase
  10.1038/s41467-022-32094-6} {\bibfield  {journal} {\bibinfo  {journal}
  {Nature Communications}\ }\textbf {\bibinfo {volume} {13}},\ \bibinfo {pages}
  {4657} (\bibinfo {year} {2022})}\BibitemShut {NoStop}%
\bibitem [{\citenamefont {Kang}\ \emph {et~al.}(2023)\citenamefont {Kang},
  \citenamefont {Campbell},\ and\ \citenamefont {Brown}}]{PRXQuantum.4.020358}%
  \BibitemOpen
  \bibfield  {author} {\bibinfo {author} {\bibfnamefont {M.}~\bibnamefont
  {Kang}}, \bibinfo {author} {\bibfnamefont {W.~C.}\ \bibnamefont {Campbell}},
  \ and\ \bibinfo {author} {\bibfnamefont {K.~R.}\ \bibnamefont {Brown}},\
  }\href {\doibase 10.1103/PRXQuantum.4.020358} {\bibfield  {journal} {\bibinfo
   {journal} {PRX Quantum}\ }\textbf {\bibinfo {volume} {4}},\ \bibinfo {pages}
  {020358} (\bibinfo {year} {2023})}\BibitemShut {NoStop}%
\bibitem [{\citenamefont {Kubica}\ \emph {et~al.}(2023)\citenamefont {Kubica},
  \citenamefont {Haim}, \citenamefont {Vaknin}, \citenamefont {Levine},
  \citenamefont {Brand\~ao},\ and\ \citenamefont
  {Retzker}}]{PhysRevX.13.041022}%
  \BibitemOpen
  \bibfield  {author} {\bibinfo {author} {\bibfnamefont {A.}~\bibnamefont
  {Kubica}}, \bibinfo {author} {\bibfnamefont {A.}~\bibnamefont {Haim}},
  \bibinfo {author} {\bibfnamefont {Y.}~\bibnamefont {Vaknin}}, \bibinfo
  {author} {\bibfnamefont {H.}~\bibnamefont {Levine}}, \bibinfo {author}
  {\bibfnamefont {F.}~\bibnamefont {Brand\~ao}}, \ and\ \bibinfo {author}
  {\bibfnamefont {A.}~\bibnamefont {Retzker}},\ }\href {\doibase
  10.1103/PhysRevX.13.041022} {\bibfield  {journal} {\bibinfo  {journal} {Phys.
  Rev. X}\ }\textbf {\bibinfo {volume} {13}},\ \bibinfo {pages} {041022}
  (\bibinfo {year} {2023})}\BibitemShut {NoStop}%
\bibitem [{\citenamefont {Levine}\ \emph {et~al.}(2023)\citenamefont {Levine}
  \emph {et~al.}}]{levine2023demonstrating}%
  \BibitemOpen
  \bibfield  {author} {\bibinfo {author} {\bibfnamefont {H.}~\bibnamefont
  {Levine}} \emph {et~al.},\ }\href@noop {} {\enquote {\bibinfo {title}
  {Demonstrating a long-coherence dual-rail erasure qubit using tunable
  transmons},}\ } (\bibinfo {year} {2023}),\ \Eprint
  {http://arxiv.org/abs/2307.08737} {arXiv:2307.08737 [quant-ph]} \BibitemShut
  {NoStop}%
\bibitem [{\citenamefont {Steane}(1997)}]{steaneec}%
  \BibitemOpen
  \bibfield  {author} {\bibinfo {author} {\bibfnamefont {A.~M.}\ \bibnamefont
  {Steane}},\ }\href {\doibase 10.1103/PhysRevLett.78.2252} {\bibfield
  {journal} {\bibinfo  {journal} {Phys. Rev. Lett.}\ }\textbf {\bibinfo
  {volume} {78}},\ \bibinfo {pages} {2252} (\bibinfo {year}
  {1997})}\BibitemShut {NoStop}%
\bibitem [{\citenamefont {Bacon}\ \emph {et~al.}(2017)\citenamefont {Bacon},
  \citenamefont {Flammia}, \citenamefont {Harrow},\ and\ \citenamefont
  {Shi}}]{bfhs}%
  \BibitemOpen
  \bibfield  {author} {\bibinfo {author} {\bibfnamefont {D.}~\bibnamefont
  {Bacon}}, \bibinfo {author} {\bibfnamefont {S.~T.}\ \bibnamefont {Flammia}},
  \bibinfo {author} {\bibfnamefont {A.~W.}\ \bibnamefont {Harrow}}, \ and\
  \bibinfo {author} {\bibfnamefont {J.}~\bibnamefont {Shi}},\ }\href {\doibase
  10.1109/tit.2017.2663199} {\bibfield  {journal} {\bibinfo  {journal} {IEEE
  Transactions on Information Theory}\ }\textbf {\bibinfo {volume} {63}},\
  \bibinfo {pages} {2464–2479} (\bibinfo {year} {2017})}\BibitemShut
  {NoStop}%
\bibitem [{\citenamefont {Delfosse}\ and\ \citenamefont
  {Paetznick}(2023)}]{delfosse2023spacetime}%
  \BibitemOpen
  \bibfield  {author} {\bibinfo {author} {\bibfnamefont {N.}~\bibnamefont
  {Delfosse}}\ and\ \bibinfo {author} {\bibfnamefont {A.}~\bibnamefont
  {Paetznick}},\ }\href@noop {} {\enquote {\bibinfo {title} {Spacetime codes of
  {C}lifford circuits},}\ } (\bibinfo {year} {2023}),\ \Eprint
  {http://arxiv.org/abs/2304.05943} {arXiv:2304.05943 [quant-ph]} \BibitemShut
  {NoStop}%
\bibitem [{\citenamefont {Kitaev}(2003)}]{Kitaev_2003}%
  \BibitemOpen
  \bibfield  {author} {\bibinfo {author} {\bibfnamefont {A.}~\bibnamefont
  {Kitaev}},\ }\href {\doibase 10.1016/s0003-4916(02)00018-0} {\bibfield
  {journal} {\bibinfo  {journal} {Annals of Physics}\ }\textbf {\bibinfo
  {volume} {303}},\ \bibinfo {pages} {2} (\bibinfo {year} {2003})}\BibitemShut
  {NoStop}%
\bibitem [{\citenamefont {Dennis}\ \emph {et~al.}(2002)\citenamefont {Dennis},
  \citenamefont {Kitaev}, \citenamefont {Landahl},\ and\ \citenamefont
  {Preskill}}]{Dennis_2002}%
  \BibitemOpen
  \bibfield  {author} {\bibinfo {author} {\bibfnamefont {E.}~\bibnamefont
  {Dennis}}, \bibinfo {author} {\bibfnamefont {A.}~\bibnamefont {Kitaev}},
  \bibinfo {author} {\bibfnamefont {A.}~\bibnamefont {Landahl}}, \ and\
  \bibinfo {author} {\bibfnamefont {J.}~\bibnamefont {Preskill}},\ }\href
  {\doibase 10.1063/1.1499754} {\bibfield  {journal} {\bibinfo  {journal}
  {Journal of Mathematical Physics}\ }\textbf {\bibinfo {volume} {43}},\
  \bibinfo {pages} {4452} (\bibinfo {year} {2002})}\BibitemShut {NoStop}%
\bibitem [{\citenamefont {Fowler}\ \emph {et~al.}(2009)\citenamefont {Fowler},
  \citenamefont {Stephens},\ and\ \citenamefont {Groszkowski}}]{Fowler2009}%
  \BibitemOpen
  \bibfield  {author} {\bibinfo {author} {\bibfnamefont {A.~G.}\ \bibnamefont
  {Fowler}}, \bibinfo {author} {\bibfnamefont {A.~M.}\ \bibnamefont
  {Stephens}}, \ and\ \bibinfo {author} {\bibfnamefont {P.}~\bibnamefont
  {Groszkowski}},\ }\href {\doibase 10.1103/PhysRevA.80.052312} {\bibfield
  {journal} {\bibinfo  {journal} {Phys. Rev. A}\ }\textbf {\bibinfo {volume}
  {80}},\ \bibinfo {pages} {052312} (\bibinfo {year} {2009})}\BibitemShut
  {NoStop}%
\bibitem [{\citenamefont {Tremblay}\ \emph {et~al.}(2022)\citenamefont
  {Tremblay}, \citenamefont {Delfosse},\ and\ \citenamefont
  {Beverland}}]{Tremblay_2022}%
  \BibitemOpen
  \bibfield  {author} {\bibinfo {author} {\bibfnamefont {M.~A.}\ \bibnamefont
  {Tremblay}}, \bibinfo {author} {\bibfnamefont {N.}~\bibnamefont {Delfosse}},
  \ and\ \bibinfo {author} {\bibfnamefont {M.~E.}\ \bibnamefont {Beverland}},\
  }\href {\doibase 10.1103/PhysRevLett.129.050504} {\bibfield  {journal}
  {\bibinfo  {journal} {Phys. Rev. Lett.}\ }\textbf {\bibinfo {volume} {129}},\
  \bibinfo {pages} {050504} (\bibinfo {year} {2022})}\BibitemShut {NoStop}%
\bibitem [{\citenamefont {Bravyi}\ \emph {et~al.}(2023)\citenamefont {Bravyi},
  \citenamefont {Cross}, \citenamefont {Gambetta}, \citenamefont {Maslov},
  \citenamefont {Rall},\ and\ \citenamefont {Yoder}}]{bravyi2023highthreshold}%
  \BibitemOpen
  \bibfield  {author} {\bibinfo {author} {\bibfnamefont {S.}~\bibnamefont
  {Bravyi}}, \bibinfo {author} {\bibfnamefont {A.~W.}\ \bibnamefont {Cross}},
  \bibinfo {author} {\bibfnamefont {J.~M.}\ \bibnamefont {Gambetta}}, \bibinfo
  {author} {\bibfnamefont {D.}~\bibnamefont {Maslov}}, \bibinfo {author}
  {\bibfnamefont {P.}~\bibnamefont {Rall}}, \ and\ \bibinfo {author}
  {\bibfnamefont {T.~J.}\ \bibnamefont {Yoder}},\ }\href@noop {} {\enquote
  {\bibinfo {title} {High-threshold and low-overhead fault-tolerant quantum
  memory},}\ } (\bibinfo {year} {2023}),\ \Eprint
  {http://arxiv.org/abs/2308.07915} {arXiv:2308.07915 [quant-ph]} \BibitemShut
  {NoStop}%
\bibitem [{\citenamefont {Xu}\ \emph {et~al.}(2023)\citenamefont {Xu} \emph
  {et~al.}}]{xu2023constantoverhead}%
  \BibitemOpen
  \bibfield  {author} {\bibinfo {author} {\bibfnamefont {Q.}~\bibnamefont {Xu}}
  \emph {et~al.},\ }\href@noop {} {\enquote {\bibinfo {title}
  {Constant-overhead fault-tolerant quantum computation with reconfigurable
  atom arrays},}\ } (\bibinfo {year} {2023}),\ \Eprint
  {http://arxiv.org/abs/2308.08648} {arXiv:2308.08648 [quant-ph]} \BibitemShut
  {NoStop}%
\bibitem [{\citenamefont {Higgott}\ and\ \citenamefont
  {Breuckmann}(2023)}]{higgott2023constructions}%
  \BibitemOpen
  \bibfield  {author} {\bibinfo {author} {\bibfnamefont {O.}~\bibnamefont
  {Higgott}}\ and\ \bibinfo {author} {\bibfnamefont {N.~P.}\ \bibnamefont
  {Breuckmann}},\ }\href@noop {} {\enquote {\bibinfo {title} {Constructions and
  performance of hyperbolic and semi-hyperbolic floquet codes},}\ } (\bibinfo
  {year} {2023}),\ \Eprint {http://arxiv.org/abs/2308.03750} {arXiv:2308.03750
  [quant-ph]} \BibitemShut {NoStop}%
\bibitem [{\citenamefont {Fahimniya}\ \emph {et~al.}(2023)\citenamefont
  {Fahimniya}, \citenamefont {Dehghani}, \citenamefont {Bharti}, \citenamefont
  {Mathew}, \citenamefont {Kollár}, \citenamefont {Gorshkov},\ and\
  \citenamefont {Gullans}}]{fahimniya2023faulttolerant}%
  \BibitemOpen
  \bibfield  {author} {\bibinfo {author} {\bibfnamefont {A.}~\bibnamefont
  {Fahimniya}}, \bibinfo {author} {\bibfnamefont {H.}~\bibnamefont {Dehghani}},
  \bibinfo {author} {\bibfnamefont {K.}~\bibnamefont {Bharti}}, \bibinfo
  {author} {\bibfnamefont {S.}~\bibnamefont {Mathew}}, \bibinfo {author}
  {\bibfnamefont {A.~J.}\ \bibnamefont {Kollár}}, \bibinfo {author}
  {\bibfnamefont {A.~V.}\ \bibnamefont {Gorshkov}}, \ and\ \bibinfo {author}
  {\bibfnamefont {M.~J.}\ \bibnamefont {Gullans}},\ }\href@noop {} {\enquote
  {\bibinfo {title} {Fault-tolerant hyperbolic floquet quantum error correcting
  codes},}\ } (\bibinfo {year} {2023}),\ \Eprint
  {http://arxiv.org/abs/2309.10033} {arXiv:2309.10033 [quant-ph]} \BibitemShut
  {NoStop}%
\bibitem [{\citenamefont {Wang}\ \emph {et~al.}(2023)\citenamefont {Wang} \emph
  {et~al.}}]{wang2023faulttolerant}%
  \BibitemOpen
  \bibfield  {author} {\bibinfo {author} {\bibfnamefont {Y.}~\bibnamefont
  {Wang}} \emph {et~al.},\ }\href@noop {} {\enquote {\bibinfo {title}
  {Fault-tolerant one-bit addition with the smallest interesting colour
  code},}\ } (\bibinfo {year} {2023}),\ \Eprint
  {http://arxiv.org/abs/2309.09893} {arXiv:2309.09893 [quant-ph]} \BibitemShut
  {NoStop}%
\bibitem [{\citenamefont {Ryan-Anderson}\ \emph {et~al.}(2022)\citenamefont
  {Ryan-Anderson} \emph {et~al.}}]{ryananderson2022implementing}%
  \BibitemOpen
  \bibfield  {author} {\bibinfo {author} {\bibfnamefont {C.}~\bibnamefont
  {Ryan-Anderson}} \emph {et~al.},\ }\href@noop {} {\enquote {\bibinfo {title}
  {Implementing fault-tolerant entangling gates on the five-qubit code and the
  color code},}\ } (\bibinfo {year} {2022}),\ \Eprint
  {http://arxiv.org/abs/2208.01863} {arXiv:2208.01863 [quant-ph]} \BibitemShut
  {NoStop}%
\bibitem [{\citenamefont {Gupta}\ \emph {et~al.}(2023)\citenamefont {Gupta}
  \emph {et~al.}}]{gupta2023encoding}%
  \BibitemOpen
  \bibfield  {author} {\bibinfo {author} {\bibfnamefont {R.~S.}\ \bibnamefont
  {Gupta}} \emph {et~al.},\ }\href@noop {} {\enquote {\bibinfo {title}
  {Encoding a magic state with beyond break-even fidelity},}\ } (\bibinfo
  {year} {2023}),\ \Eprint {http://arxiv.org/abs/2305.13581} {arXiv:2305.13581
  [quant-ph]} \BibitemShut {NoStop}%
\bibitem [{\citenamefont {Krinner}\ \emph {et~al.}(2022)\citenamefont {Krinner}
  \emph {et~al.}}]{Krinner_2022}%
  \BibitemOpen
  \bibfield  {author} {\bibinfo {author} {\bibfnamefont {S.}~\bibnamefont
  {Krinner}} \emph {et~al.},\ }\href {\doibase 10.1038/s41586-022-04566-8}
  {\bibfield  {journal} {\bibinfo  {journal} {Nature}\ }\textbf {\bibinfo
  {volume} {605}},\ \bibinfo {pages} {669} (\bibinfo {year}
  {2022})}\BibitemShut {NoStop}%
\bibitem [{\citenamefont {Acharya}\ \emph {et~al.}(2022)\citenamefont {Acharya}
  \emph {et~al.}}]{acharya2022suppressing}%
  \BibitemOpen
  \bibfield  {author} {\bibinfo {author} {\bibfnamefont {R.}~\bibnamefont
  {Acharya}} \emph {et~al.},\ }\href@noop {} {\enquote {\bibinfo {title}
  {Suppressing quantum errors by scaling a surface code logical qubit},}\ }
  (\bibinfo {year} {2022}),\ \Eprint {http://arxiv.org/abs/2207.06431}
  {arXiv:2207.06431 [quant-ph]} \BibitemShut {NoStop}%
\bibitem [{\citenamefont {Iyer}\ and\ \citenamefont
  {Poulin}(2015)}]{iyer2015hardness}%
  \BibitemOpen
  \bibfield  {author} {\bibinfo {author} {\bibfnamefont {P.}~\bibnamefont
  {Iyer}}\ and\ \bibinfo {author} {\bibfnamefont {D.}~\bibnamefont {Poulin}},\
  }\href@noop {} {\bibfield  {journal} {\bibinfo  {journal} {IEEE Transactions
  on Information Theory}\ }\textbf {\bibinfo {volume} {61}},\ \bibinfo {pages}
  {5209} (\bibinfo {year} {2015})}\BibitemShut {NoStop}%
\bibitem [{\citenamefont {Duclos-Cianci}\ and\ \citenamefont
  {Poulin}(2010)}]{duclos2010fast}%
  \BibitemOpen
  \bibfield  {author} {\bibinfo {author} {\bibfnamefont {G.}~\bibnamefont
  {Duclos-Cianci}}\ and\ \bibinfo {author} {\bibfnamefont {D.}~\bibnamefont
  {Poulin}},\ }\href {\doibase 10.1103/PhysRevLett.104.050504} {\bibfield
  {journal} {\bibinfo  {journal} {Phys. Rev. Lett.}\ }\textbf {\bibinfo
  {volume} {104}},\ \bibinfo {pages} {050504} (\bibinfo {year}
  {2010})}\BibitemShut {NoStop}%
\bibitem [{\citenamefont {Liu}\ \emph {et~al.}(2021)\citenamefont {Liu},
  \citenamefont {Wang},\ and\ \citenamefont {Zhang}}]{liu2021tropical}%
  \BibitemOpen
  \bibfield  {author} {\bibinfo {author} {\bibfnamefont {J.-G.}\ \bibnamefont
  {Liu}}, \bibinfo {author} {\bibfnamefont {L.}~\bibnamefont {Wang}}, \ and\
  \bibinfo {author} {\bibfnamefont {P.}~\bibnamefont {Zhang}},\ }\href
  {\doibase 10.1103/PhysRevLett.126.090506} {\bibfield  {journal} {\bibinfo
  {journal} {Phys. Rev. Lett.}\ }\textbf {\bibinfo {volume} {126}},\ \bibinfo
  {pages} {090506} (\bibinfo {year} {2021})}\BibitemShut {NoStop}%
\bibitem [{\citenamefont {Chubb}\ and\ \citenamefont
  {Flammia}(2021)}]{chubb2021statistical}%
  \BibitemOpen
  \bibfield  {author} {\bibinfo {author} {\bibfnamefont {C.~T.}\ \bibnamefont
  {Chubb}}\ and\ \bibinfo {author} {\bibfnamefont {S.~T.}\ \bibnamefont
  {Flammia}},\ }\href@noop {} {\bibfield  {journal} {\bibinfo  {journal}
  {Annales de l’Institut Henri Poincar{\'e} D}\ }\textbf {\bibinfo {volume}
  {8}},\ \bibinfo {pages} {269} (\bibinfo {year} {2021})}\BibitemShut {NoStop}%
\bibitem [{\citenamefont {Nishimori}(1981)}]{nishimori1981internal}%
  \BibitemOpen
  \bibfield  {author} {\bibinfo {author} {\bibfnamefont {H.}~\bibnamefont
  {Nishimori}},\ }\href {\doibase 10.1143/PTP.66.1169} {\bibfield  {journal}
  {\bibinfo  {journal} {Progress of Theoretical Physics}\ }\textbf {\bibinfo
  {volume} {66}},\ \bibinfo {pages} {1169} (\bibinfo {year} {1981})},\ \Eprint
  {http://arxiv.org/abs/https://academic.oup.com/ptp/article-pdf/66/4/1169/5265369/66-4-1169.pdf}
  {https://academic.oup.com/ptp/article-pdf/66/4/1169/5265369/66-4-1169.pdf}
  \BibitemShut {NoStop}%
\bibitem [{\citenamefont {Aaronson}\ and\ \citenamefont
  {Gottesman}(2004)}]{Aaronson_2004}%
  \BibitemOpen
  \bibfield  {author} {\bibinfo {author} {\bibfnamefont {S.}~\bibnamefont
  {Aaronson}}\ and\ \bibinfo {author} {\bibfnamefont {D.}~\bibnamefont
  {Gottesman}},\ }\href {\doibase 10.1103/physreva.70.052328} {\bibfield
  {journal} {\bibinfo  {journal} {Physical Review A}\ }\textbf {\bibinfo
  {volume} {70}},\ \bibinfo {pages} {052328} (\bibinfo {year}
  {2004})}\BibitemShut {NoStop}%
\bibitem [{Note1()}]{Note1}%
  \BibitemOpen
  \bibinfo {note} {Note that $\protect \bar {p}_L$ is similar to $p'_L$ in
  \cite {darmawan2022low} in that they both consider failure probabilities of
  individual logical qubits.}\BibitemShut {Stop}%
\bibitem [{\citenamefont {Wang}\ \emph {et~al.}(2003)\citenamefont {Wang},
  \citenamefont {Harrington},\ and\ \citenamefont {Preskill}}]{Wang_2003}%
  \BibitemOpen
  \bibfield  {author} {\bibinfo {author} {\bibfnamefont {C.}~\bibnamefont
  {Wang}}, \bibinfo {author} {\bibfnamefont {J.}~\bibnamefont {Harrington}}, \
  and\ \bibinfo {author} {\bibfnamefont {J.}~\bibnamefont {Preskill}},\ }\href
  {\doibase 10.1016/s0003-4916(02)00019-2} {\bibfield  {journal} {\bibinfo
  {journal} {Annals of Physics}\ }\textbf {\bibinfo {volume} {303}},\ \bibinfo
  {pages} {31} (\bibinfo {year} {2003})}\BibitemShut {NoStop}%
\bibitem [{\citenamefont {Harrington}(2004)}]{Harrington2004AnalysisOQ}%
  \BibitemOpen
  \bibfield  {author} {\bibinfo {author} {\bibfnamefont {J.~W.}\ \bibnamefont
  {Harrington}}\ }(\bibinfo {year} {2004})\BibitemShut {NoStop}%
\bibitem [{\citenamefont {Stace}\ and\ \citenamefont
  {Barrett}(2010)}]{PhysRevA.81.022317}%
  \BibitemOpen
  \bibfield  {author} {\bibinfo {author} {\bibfnamefont {T.~M.}\ \bibnamefont
  {Stace}}\ and\ \bibinfo {author} {\bibfnamefont {S.~D.}\ \bibnamefont
  {Barrett}},\ }\href {\doibase 10.1103/PhysRevA.81.022317} {\bibfield
  {journal} {\bibinfo  {journal} {Phys. Rev. A}\ }\textbf {\bibinfo {volume}
  {81}},\ \bibinfo {pages} {022317} (\bibinfo {year} {2010})}\BibitemShut
  {NoStop}%
\bibitem [{\citenamefont {Chamberland}\ \emph {et~al.}(2017)\citenamefont
  {Chamberland}, \citenamefont {Jochym-O’Connor},\ and\ \citenamefont
  {Laflamme}}]{Chamberland_2017}%
  \BibitemOpen
  \bibfield  {author} {\bibinfo {author} {\bibfnamefont {C.}~\bibnamefont
  {Chamberland}}, \bibinfo {author} {\bibfnamefont {T.}~\bibnamefont
  {Jochym-O’Connor}}, \ and\ \bibinfo {author} {\bibfnamefont
  {R.}~\bibnamefont {Laflamme}},\ }\href {\doibase 10.1103/physreva.95.022313}
  {\bibfield  {journal} {\bibinfo  {journal} {Physical Review A}\ }\textbf
  {\bibinfo {volume} {95}} (\bibinfo {year} {2017}),\
  10.1103/physreva.95.022313}\BibitemShut {NoStop}%
\bibitem [{\citenamefont {Nahum}\ \emph {et~al.}(2017)\citenamefont {Nahum},
  \citenamefont {Ruhman}, \citenamefont {Vijay},\ and\ \citenamefont
  {Haah}}]{PhysRevX.7.031016}%
  \BibitemOpen
  \bibfield  {author} {\bibinfo {author} {\bibfnamefont {A.}~\bibnamefont
  {Nahum}}, \bibinfo {author} {\bibfnamefont {J.}~\bibnamefont {Ruhman}},
  \bibinfo {author} {\bibfnamefont {S.}~\bibnamefont {Vijay}}, \ and\ \bibinfo
  {author} {\bibfnamefont {J.}~\bibnamefont {Haah}},\ }\href {\doibase
  10.1103/PhysRevX.7.031016} {\bibfield  {journal} {\bibinfo  {journal} {Phys.
  Rev. X}\ }\textbf {\bibinfo {volume} {7}},\ \bibinfo {pages} {031016}
  (\bibinfo {year} {2017})}\BibitemShut {NoStop}%
\bibitem [{\citenamefont {Hamma}\ \emph
  {et~al.}(2005{\natexlab{a}})\citenamefont {Hamma}, \citenamefont
  {Ionicioiu},\ and\ \citenamefont {Zanardi}}]{PhysRevA.71.022315}%
  \BibitemOpen
  \bibfield  {author} {\bibinfo {author} {\bibfnamefont {A.}~\bibnamefont
  {Hamma}}, \bibinfo {author} {\bibfnamefont {R.}~\bibnamefont {Ionicioiu}}, \
  and\ \bibinfo {author} {\bibfnamefont {P.}~\bibnamefont {Zanardi}},\ }\href
  {\doibase 10.1103/PhysRevA.71.022315} {\bibfield  {journal} {\bibinfo
  {journal} {Phys. Rev. A}\ }\textbf {\bibinfo {volume} {71}},\ \bibinfo
  {pages} {022315} (\bibinfo {year} {2005}{\natexlab{a}})}\BibitemShut
  {NoStop}%
\bibitem [{\citenamefont {Hamma}\ \emph
  {et~al.}(2005{\natexlab{b}})\citenamefont {Hamma}, \citenamefont
  {Ionicioiu},\ and\ \citenamefont {Zanardi}}]{HAMMA200522}%
  \BibitemOpen
  \bibfield  {author} {\bibinfo {author} {\bibfnamefont {A.}~\bibnamefont
  {Hamma}}, \bibinfo {author} {\bibfnamefont {R.}~\bibnamefont {Ionicioiu}}, \
  and\ \bibinfo {author} {\bibfnamefont {P.}~\bibnamefont {Zanardi}},\ }\href
  {\doibase https://doi.org/10.1016/j.physleta.2005.01.060} {\bibfield
  {journal} {\bibinfo  {journal} {Physics Letters A}\ }\textbf {\bibinfo
  {volume} {337}},\ \bibinfo {pages} {22} (\bibinfo {year}
  {2005}{\natexlab{b}})}\BibitemShut {NoStop}%
\bibitem [{\citenamefont {Delfosse}\ and\ \citenamefont
  {Z\'emor}(2020{\natexlab{b}})}]{delfossezemor2020}%
  \BibitemOpen
  \bibfield  {author} {\bibinfo {author} {\bibfnamefont {N.}~\bibnamefont
  {Delfosse}}\ and\ \bibinfo {author} {\bibfnamefont {G.}~\bibnamefont
  {Z\'emor}},\ }\href {\doibase 10.1103/PhysRevResearch.2.033042} {\bibfield
  {journal} {\bibinfo  {journal} {Phys. Rev. Res.}\ }\textbf {\bibinfo {volume}
  {2}},\ \bibinfo {pages} {033042} (\bibinfo {year}
  {2020}{\natexlab{b}})}\BibitemShut {NoStop}%
\bibitem [{Som()}]{Sommers_2023_trees}%
  \BibitemOpen
  \href@noop {} {}\bibinfo {note} {G. Sommers, D. A. Huse, and M. J. Gullans,
  in preparation.}\BibitemShut {Stop}%
\bibitem [{\citenamefont {Li}\ \emph {et~al.}(2018)\citenamefont {Li},
  \citenamefont {Chen},\ and\ \citenamefont {Fisher}}]{li2018quantum}%
  \BibitemOpen
  \bibfield  {author} {\bibinfo {author} {\bibfnamefont {Y.}~\bibnamefont
  {Li}}, \bibinfo {author} {\bibfnamefont {X.}~\bibnamefont {Chen}}, \ and\
  \bibinfo {author} {\bibfnamefont {M.~P.~A.}\ \bibnamefont {Fisher}},\ }\href
  {\doibase 10.1103/PhysRevB.98.205136} {\bibfield  {journal} {\bibinfo
  {journal} {Phys. Rev. B}\ }\textbf {\bibinfo {volume} {98}},\ \bibinfo
  {pages} {205136} (\bibinfo {year} {2018})}\BibitemShut {NoStop}%
\bibitem [{\citenamefont {Skinner}\ \emph {et~al.}(2019)\citenamefont
  {Skinner}, \citenamefont {Ruhman},\ and\ \citenamefont
  {Nahum}}]{skinner2019measurement}%
  \BibitemOpen
  \bibfield  {author} {\bibinfo {author} {\bibfnamefont {B.}~\bibnamefont
  {Skinner}}, \bibinfo {author} {\bibfnamefont {J.}~\bibnamefont {Ruhman}}, \
  and\ \bibinfo {author} {\bibfnamefont {A.}~\bibnamefont {Nahum}},\ }\href
  {\doibase 10.1103/PhysRevX.9.031009} {\bibfield  {journal} {\bibinfo
  {journal} {Phys. Rev. X}\ }\textbf {\bibinfo {volume} {9}},\ \bibinfo {pages}
  {031009} (\bibinfo {year} {2019})}\BibitemShut {NoStop}%
\bibitem [{\citenamefont {Chan}\ \emph {et~al.}(2019)\citenamefont {Chan},
  \citenamefont {Nandkishore}, \citenamefont {Pretko},\ and\ \citenamefont
  {Smith}}]{chan2019unitary}%
  \BibitemOpen
  \bibfield  {author} {\bibinfo {author} {\bibfnamefont {A.}~\bibnamefont
  {Chan}}, \bibinfo {author} {\bibfnamefont {R.~M.}\ \bibnamefont
  {Nandkishore}}, \bibinfo {author} {\bibfnamefont {M.}~\bibnamefont {Pretko}},
  \ and\ \bibinfo {author} {\bibfnamefont {G.}~\bibnamefont {Smith}},\ }\href
  {\doibase 10.1103/PhysRevB.99.224307} {\bibfield  {journal} {\bibinfo
  {journal} {Phys. Rev. B}\ }\textbf {\bibinfo {volume} {99}},\ \bibinfo
  {pages} {224307} (\bibinfo {year} {2019})}\BibitemShut {NoStop}%
\bibitem [{\citenamefont {Sommers}\ \emph {et~al.}(2022)\citenamefont
  {Sommers}, \citenamefont {Huse},\ and\ \citenamefont
  {Gullans}}]{sommers2022crystalline}%
  \BibitemOpen
  \bibfield  {author} {\bibinfo {author} {\bibfnamefont {G.~M.}\ \bibnamefont
  {Sommers}}, \bibinfo {author} {\bibfnamefont {D.~A.}\ \bibnamefont {Huse}}, \
  and\ \bibinfo {author} {\bibfnamefont {M.~J.}\ \bibnamefont {Gullans}},\
  }\href {\doibase 10.48550/ARXIV.2210.10808} {\enquote {\bibinfo {title}
  {Crystalline quantum circuits},}\ } (\bibinfo {year} {2022})\BibitemShut
  {NoStop}%
\end{thebibliography}%

\section{Supplementary Figures}
\label{app:supplresults}

Here, we present the crossing plots and associated collapses for the code capacity and fault-tolerant threshold experiments. Results for encoding rates of $1/3$, $1/4$, $1/5$, $1/10$, and $1/50$ are shown.
\begin{figure*}
    
    \includegraphics[width=0.6\linewidth]{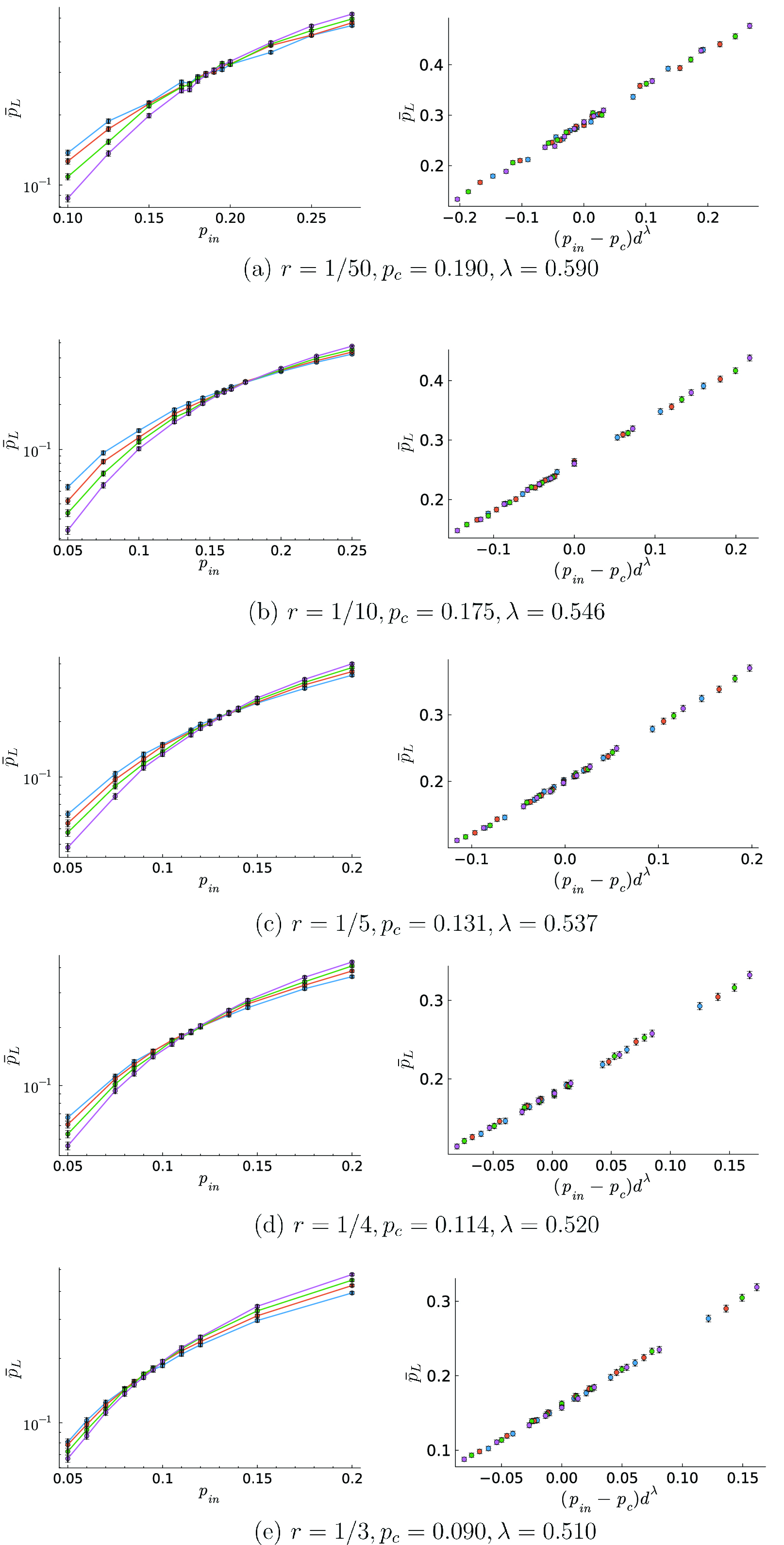}
        \caption{Code capacity threshold estimation for random-circuit codes using marginal decoding. Codes are generated on systems of $N = 50$ qubits (excluding the boundary padding) using random stabilizer circuits of depths $d = 4,5,6,7$ (blue, orange, green, purple). The crossing point signifies the threshold of the code and is estimated using our finite-size scaling ansatz, which is displayed to the right of the crossing plots. Each datapoint is estimated by sampling 10,000 random codes and Pauli errors.
}
        \label{fig:app_marg}
\end{figure*}
\begin{figure*}
    
    \includegraphics[width=0.6\linewidth]{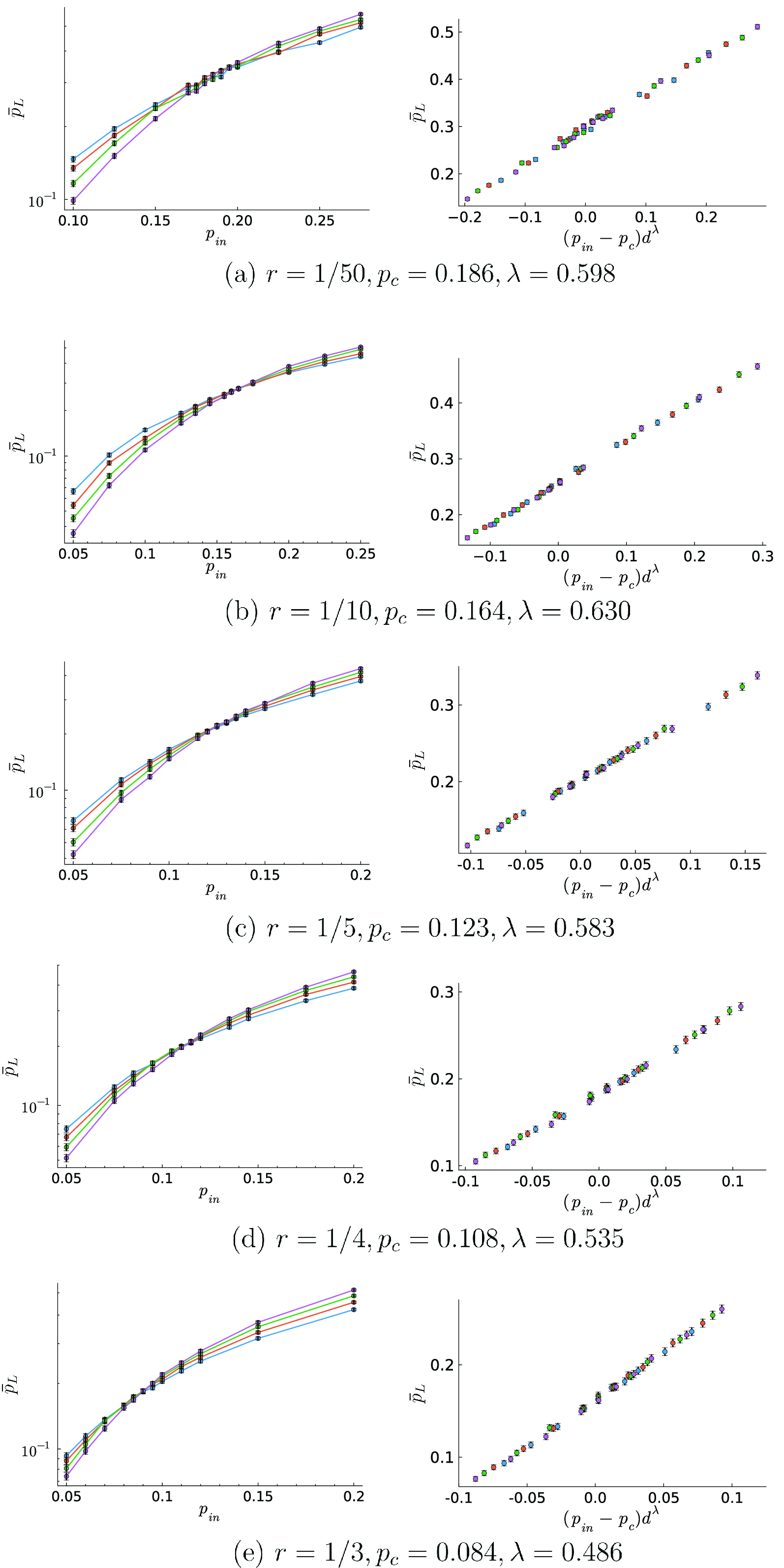}
        \caption{Code capacity threshold estimation for random-circuit codes using minimum-weight decoding. Codes are generated on systems of $N = 50$ qubits (excluding the boundary padding) using random stabilizer circuits of depths $d = 4,5,6,7$ (blue, orange, green, purple). The threshold is estimated as in Figure \ref{fig:app_marg}.}
        \label{fig:app_mw}
\end{figure*}
\begin{figure*}
    
    \includegraphics[width=0.6\linewidth]{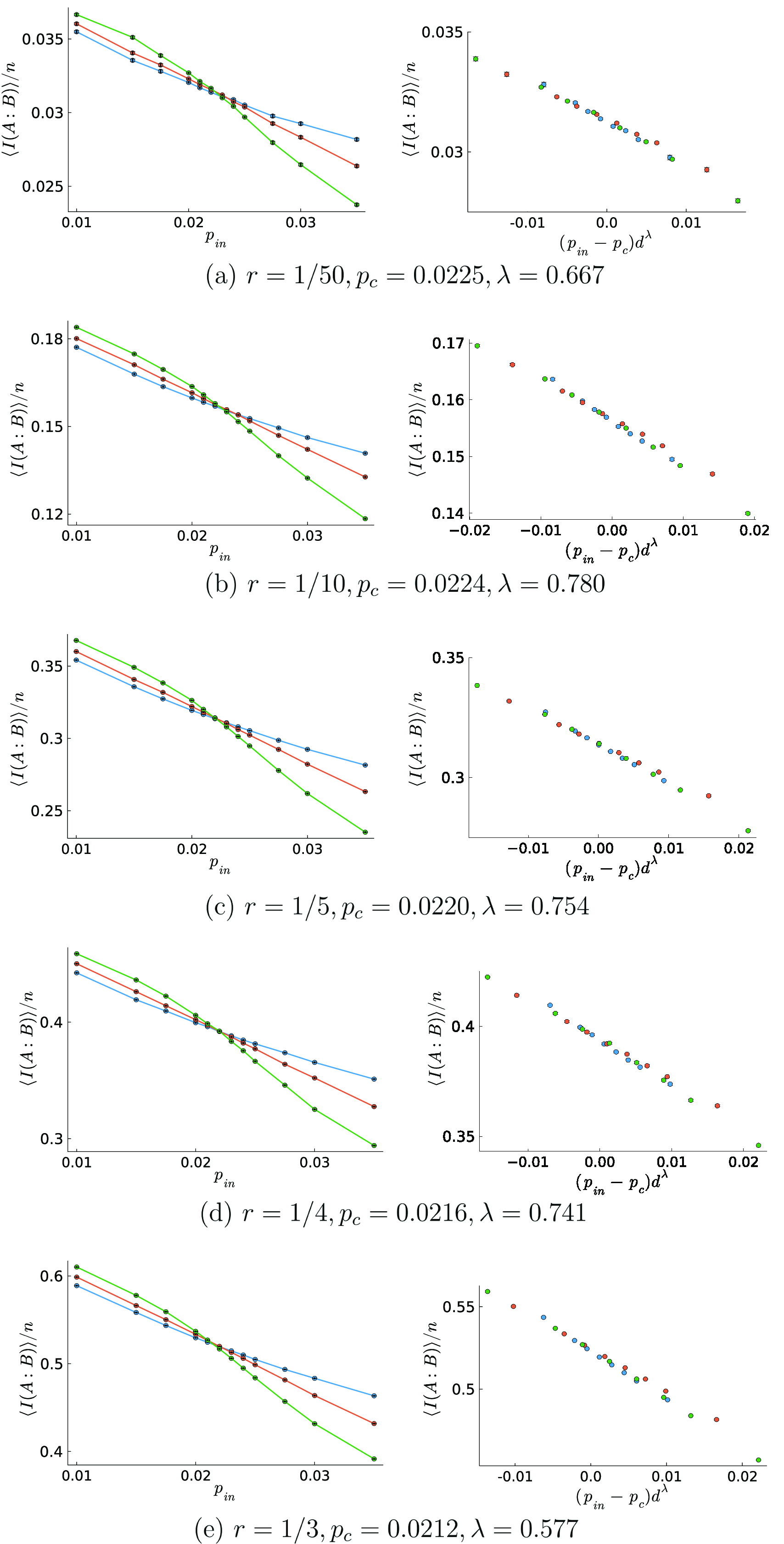}
        \caption{Fault-tolerant threshold estimation for random CSS codes at various encoding rates. Results are shown for $d=2$, $q=2$ (blue), $d=4$, $q=4$ (orange), and $d=6$, $q=6$ (green). The finite-size scaling collapse for the estimated threshold and ansatz parameters is presented on the right. $13,000$ samples are taken for the points closest to the crossing point.}
        \label{fig:app_mi}
\end{figure*}
\end{document}